%% file: main.tex
\RequirePackage[svgnames]{xcolor}

\documentclass[11pt,letterpaper]{mystyle}

\usepackage[all]{hypcap}
\usepackage[svgnames]{xcolor}
\usepackage[comma,authoryear,compress]{natbib}
\usepackage{hyperref}[citecolor=lightblue]

\hypersetup{
    colorlinks = true,
    citecolor = {YaleBlue},
}

\usepackage{algorithm}
\usepackage{algorithmicx}
\usepackage{algpseudocode}
\usepackage{microtype}
\usepackage{graphicx}
\expandafter\def\csname ver@subfig.sty\endcsname{}
\usepackage{booktabs} %
\usepackage{float}
\usepackage{bigstrut}

\usepackage{amsmath}
\usepackage{amssymb}
\usepackage{mathtools}
\usepackage{amsthm}
\usepackage{mathrsfs}
\usepackage{nicefrac}
\usepackage{dsfont}
\usepackage{enumitem}
\usepackage{subcaption}
\usepackage{graphicx,subfig}
\usepackage{cleveref}
\usepackage{bxcoloremoji}

\usepackage{float}

\usepackage[utf8]{inputenc} %
\usepackage[T1]{fontenc}    %
\usepackage{hyperref}       %
\usepackage{url}            %
\usepackage{booktabs}       %
\usepackage{amsfonts}       %
\usepackage{nicefrac}       %
\usepackage{microtype}      %
\usepackage{graphicx}
\usepackage{subcaption} 
\usepackage{amssymb}
\usepackage{fdsymbol}
\usepackage{wrapfig}
\usepackage{lipsum}
\usepackage{enumitem}
\usepackage{stackengine}
\usepackage[font=small,labelfont=bf]{caption}
\usepackage{color}
\usepackage{adjustbox}

\usepackage{rotating}
\usepackage{makecell}

\input{macro}

\newtcolorbox{AIbox}[2][]{aibox,title=#2,#1}
\definecolor{lightblue}{rgb}{0.22,0.45,0.70}%
\definecolor{Gray}{gray}{0.95}
\definecolor{Cornsilk}{rgb}{1.0, 0.97, 0.86}

\usepackage{amsmath}

\usepackage[all]{hypcap}

\newcommand{\dataset}{\textit{LLMSCBench}\xspace}

\usepackage{xspace}
\usepackage{pifont}

\title{\paperlogo{} Understanding the Supply Chain and Risks of Large Language Model Applications}

\runningtitle{\paperlogo{} Understanding the Supply Chain and Risks of Large Language Model Applications}

\author[1]{Yujie Ma}
\author[1]{Lili Quan}
\author[2]{Xiaofei Xie}
\author[1]{Qiang Hu}
\author[2]{Jiongchi Yu}
\author[1]{Yao Zhang}
\author[3]{Sen Chen}

\affil[1]{Tianjin University}
\affil[2]{Singapore Management University}
\affil[3]{Nankai University}

\begin{document}

\input{sections/abstract}

\maketitle
\vspace{3mm}

\input{sections/content}

\clearpage
\bibliography{main}

\end{document}

%% file: macro.tex
\newcommand{\x}{\mathbf{x}}

\definecolor{blanchedalmond}{rgb}{1.0, 0.92, 0.8}
\definecolor{carmine}{rgb}{0.59, 0.0, 0.09}
\definecolor{lightblue}{rgb}{0.22,0.45,0.70}%

\renewcommand{\mathbf}{\boldsymbol}

\makeatletter
\def\Ddots{\mathinner{\mkern1mu\raise\p@
\vbox{\kern7\p@\hbox{.}}\mkern2mu
\raise4\p@\hbox{.}\mkern2mu\raise7\p@\hbox{.}\mkern1mu}}
\makeatother

\definecolor{amaranth}{rgb}{0.9, 0.17, 0.31}
\definecolor{antiquebrass}{rgb}{0.8, 0.58, 0.46}
\definecolor{antiquefuchsia}{rgb}{0.57, 0.36, 0.51}
\definecolor{chromeyellow}{rgb}{0.31, 0.47, 0.26}

\newcommand{\paperlogo}{\raisebox{-1pt}{\includegraphics[height=3em]{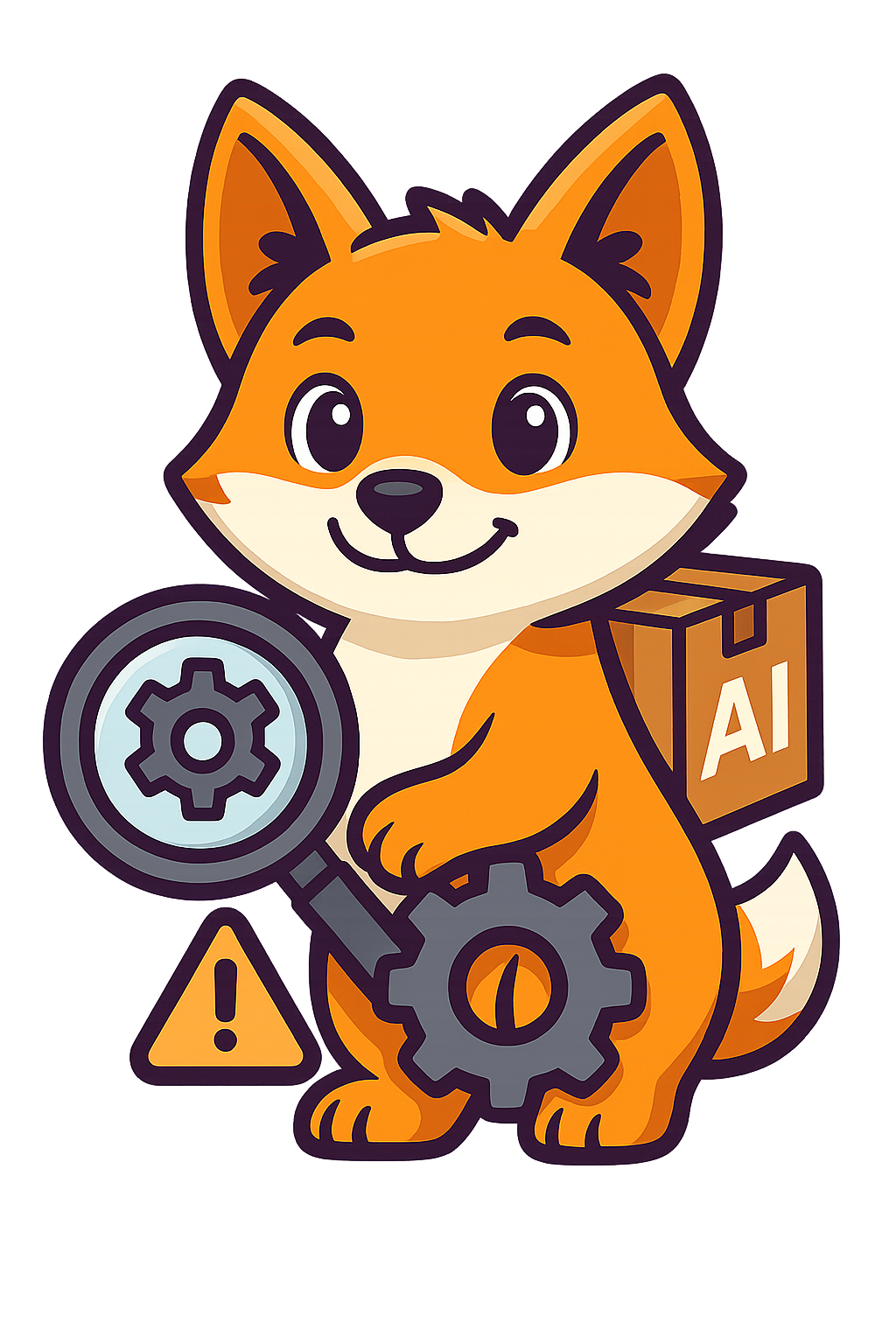}}}

%% file: sections/abstract.tex
\begin{abstract}
Large Language Models~(LLMs) have seen rapid development, leading to the widespread deployment of LLM-based systems across various domains. As these systems become increasingly applied into real-world scenarios, understanding the potential risks associated with their complex ecosystems is of growing importance. LLM-based systems are not standalone models--they rely on intricate supply chains comprising pretrained models, third-party libraries, datasets, and infrastructure components. However, most existing risk assessments focus narrowly on the model level or data level, overlooking the broader supply chain context. While recent studies have begun exploring LLM supply chain risks, there is still a lack of benchmarks to support systematic research in this area.

To bridge this gap, we introduce \dataset, the first comprehensive dataset aimed at facilitating the analysis and benchmarking of LLM supply chain security. We begin by collecting a large set of real-world LLM applications, and perform inter-dependency analysis to identify the core components used—including models, datasets, and implementation libraries. We then perform the supply chain dependency analysis to extract model fine-tuning paths, dataset reuse, and library reliance. In total, \dataset comprises 3,859 LLM applications, 109,211 models, 2,474 datasets, and 8,862 libraries, along with their dependency relationships.
To assess the security posture of the LLM ecosystem, we extract a total of 1,555 risk-related issues: 50 for applications, 325 for models, 18 for datasets, and 1,229 for libraries, sourced from various public vulnerability databases and repositories. 

Using the constructed dataset, we conduct an empirical study to analyze the dependencies among different components and their associated risks. Our findings reveal that LLM-based applications involve complex and deeply nested dependencies, highlighting the critical need for comprehensive supply chain analysis. Furthermore, our risk analysis uncovers significant and widespread vulnerabilities across the LLM supply chain, emphasizing the urgency of adopting more robust security and privacy practices. We conclude with practical implications and actionable recommendations for researchers and developers to guide the development of safe, secure, and trustworthy LLM-enabled systems.

\vspace{2mm}

\textit{Keywords: Risk Analysis, Large Language Model, Software Supply Chain}









\end{abstract}

%% file: sections/content.tex
\section{Introduction}
\label{sec:introduction}

The rapid development of large language models (LLMs) with their impressive results in various fields brings hope to general artificial intelligence. As LLM-based applications increase significantly, with their important role in our daily lives, LLM system risk management becomes increasingly important. Recent incidents have highlighted serious risks in the LLM ecosystem that can lead to significant consequences. For example, poisoning attacks on open-source models like DeepSeek can cause the generation of toxic outputs~\cite{deepseek-issue}, while malicious LLMs uploaded to public model hubs have successfully bypassed security scans, threatening downstream applications~\cite{zhu2025my}. These emerging threats underscore the urgency of ensuring the security and trustworthiness of LLM-based applications.    

Although extensive research has been conducted on LLM security, most existing efforts focus primarily on the model itself. For example, numerous studies have explored jailbreak attacks~\cite{xu2024comprehensive,wei2023jailbroken,shen2024anything} and adversarial attacks~\cite{carlini2023aligned,liu2024exploring,shayegani2023survey,zou2023universal} to assess the robustness of LLMs. However, LLM-based applications are complex systems, and while the model is a central component, it is far from the only one. These applications also rely on various other elements, including datasets~\cite{chang2024survey}, third-party libraries~\cite{TensorFlowCVE,253266,models-are-code}, and supporting application code~\cite{su2024gpt,hui2024pleak,liu2023prompt,liu2310prompt,pedro2023prompt}. Risks associated with these non-model components can also critically impact the security and trustworthiness of the final deployed applications. Therefore, a comprehensive investigation that accounts for the risks across all components, as well as their interdependent supply chains, is essential for securing LLM-based systems in real-world settings.

Recently, the software engineering community has begun to examine the complexity of LLM-based systems and their associated supply chains (LLMSC)~\cite{wang2025large,huang2024lifting,gunay2025review,hu2025large}. These studies have systematically analyzed potential risks in LLMSC from various perspectives, including programs, models, and datasets, highlighting the importance of safeguarding all components within LLM ecosystems. However, while these works underscore the urgent need for system-level analysis of LLM supply chain risks, they remain largely conceptual and high-level analysis. To enable in-depth empirical analysis and support the development of practical mitigation techniques, there is still a critical lack of comprehensive benchmark datasets specifically designed for LLMSC.

To address this gap, we construct the first benchmark dataset, \dataset, to support comprehensive risk analysis of LLM supply chains. \dataset includes a wide range of popular LLM applications, along with the models, datasets, and implementation libraries they utilize. It also captures the dependencies among these components and their associated potential risk issues. To build \dataset, we begin by extracting LLM-integrated applications from the widely used Hugging Face platform~\cite{huggingface-Spaces}. We then perform an inter-dependency analysis to identify the specific LLMs, datasets, and libraries used in each application. To gain a deeper understanding of the broader supply chain risks, we further analyze the dependencies of these components, i.e., the additional models, datasets, and libraries that they rely on. For clarity, throughout the rest of the paper, we refer to the components directly included in an application as \textit{base components} (i.e., base models, base datasets, and base libraries), and their dependencies as \textit{dependent components} (i.e., dependent models, dependent datasets, and dependent libraries). In total, \dataset covers 3,859 LLM applications, 109,211 models, 2,474 datasets, and 8,862 libraries, including both base and dependent components.

To assess the potential risks within the LLM supply chain, we further collect and analyze risk-related issues associated with each component and the applications as a whole. Following prior studies~\cite{slattery2024ai, das2025security}, we define 13 categories of risks, such as toxicity, misinformation, and malicious actors, to guide our analysis. Using these risk types as keywords, we crawl and extract user-reported issues for applications, models, and datasets from relevant platforms. For libraries, we directly obtain vulnerability information by collecting their corresponding Common Vulnerabilities and Exposures (CVE) entries from OSV.dev~\cite{osv-dev}. In total, \dataset includes 50, 325, 18, and 1,229 risk issues associated with applications, models, datasets, and libraries, respectively. 

Based on the constructed dataset, we conduct a systematic exploratory study to examine the structural complexity of LLM applications, their dependencies (e.g., the popularity and reuse of specific LLMs), and the distribution of risk issues across the LLM supply chain (LLMSC). Our analysis reveals several key insights. 
First, LLM applications often exhibit deep and diverse dependency chains, with popular models and libraries being extensively reused or dependent across multiple applications, for example, 42.8\% of models rely on other models. In terms of risk, we categorize the types of issues observed across different components and highlight the most prevalent threats. Some applications, such as \textit{text-summarizer-for-news-articles}, depend on more than 20 libraries, many of which contain critical CVEs. Moreover, these risks have a high potential for propagation, posing threats not only to the immediate application but also to downstream components and dependent systems. These findings underscore the urgent need for robust supply chain risk analysis and effective dependency management practices. Based on our results, we offer concrete implications for developers, researchers, and policymakers to guide future efforts in securing LLM systems and enabling more trustworthy supply chain ecosystems.

In summary, this paper makes the following contributions:

\begin{itemize}[leftmargin=*]
    \item We propose \dataset, the first benchmark dataset that covers LLM applications, models, datasets, and libraries, accompanied with their risk issues, to support the risk assurance research in the domain of LLMSC. 

    \item We construct detailed dependency relationships among the base components, enabling fine-grained supply chain analysis and facilitating the study of risk propagation across the LLMSC.

    \item Based on \dataset, we conduct a systematic exploratory study to investigate the dependency structures and risk characteristics of each component within the LLMSC, and provide actionable insights and implications for future research and the secure development of LLM-enabled applications.

\end{itemize}

The construction of \dataset represents a critical step toward advancing the understanding and secure deployment of complex LLM-based systems. By capturing the full ecosystem, including applications, models, datasets, libraries, and their intricate dependency relationships, \dataset enables a holistic view of LLM supply chains. This comprehensive perspective is essential for identifying not only direct risks, but also potential transitive vulnerabilities and attack propagation paths that can arise from interconnected components. As the adoption of LLMs continues to grow, we believe \dataset will serve as a timely and indispensable resource for both the research community and industry practitioners.

\section{Background}
\label{sec:background}

\begin{figure}[t]
	\centering
	\includegraphics[width=.8\textwidth]{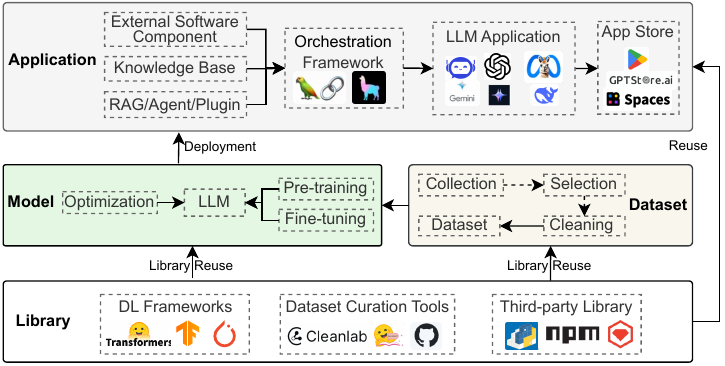}
	\caption{Components in the LLM supply chain.}
	\label{fig:LLMSC}
\end{figure}

LLMSC encompasses the entire lifecycle of LLMs \cite{wang2025large}, involving multiple stages such as dataset creation and preprocessing, model training, model optimization, and model deployment. As shown in \Cref{fig:LLMSC}, the LLMSC includes four core components: library, dataset, model, and application, which collectively affect the performance and risks of deploying LLMs in real-world applications.

\subsection{Large Language Model Supply Chain}

First, the library encompasses all toolchains and dependencies required throughout LLM development, deployment, and application, such as DL frameworks, data processing tools, and various third-party libraries. Second, the dataset includes large-scale text corpora as well as specialized domain datasets used for LLM pre-training and fine-tuning. Third, models refer to the core artifacts of LLM systems, including trained model weights, configuration files, and various model formats, which are produced and refined through stages such as architecture design, pre-training, fine-tuning, testing, release, deployment, and maintenance. Finally, the application component enables the integration of LLMs into diverse intelligent systems such as chatbots and domain-specific LLM solutions.

\subsection{Security Issues in Large Language Models}

Within the LLM supply chain, core components are interconnected and serve as sources of security issues for LLM systems. Specifically, insecure libraries can enable attackers to compromise the entire LLM pipeline; for example, TensorFlow’s Keras framework has experienced vulnerabilities such as CVE-2024-3660, which allowed arbitrary code injection \cite{TensorFlowCVE,253266,models-are-code}, while PyTorch’s use of Pickle for model serialization introduces potential deserialization risks. For datasets, previous studies \cite{chang2024survey} have highlighted a range of threats, including backdoor and poisoning attacks, privacy leakage, unauthorized disclosure, data bias, toxicity, and various ethical concerns. Some of these risks, such as data poisoning, can directly degrade model performance or inject backdoors, undermining both prediction integrity and user security.

Furthermore, many studies \cite{cui2024risk, gehman2020realtoxicityprompts,huang2023flames} have shown that LLMs may exhibit harmful biases, hallucinations, stereotypes, or discriminatory behavior. In addition, they are vulnerable to adversarial attacks \cite{carlini2023aligned, liu2024exploring, shayegani2023survey, zou2023universal}, jailbreaks \cite{xu2024comprehensive, wei2023jailbroken, shen2024anything}, backdoor attacks \cite{li2024backdoorllm, zhao2024survey}, and misuse \cite{bhatt2023purple}, all of which can lead to harmful outputs. At the application layer, LLM-powered applications face significant challenges, particularly in the context of prompt injection attacks \cite{liu2023prompt,liu2310prompt,pedro2023prompt}, misuse of function calling features, and prompt leakage \cite{su2024gpt,hui2024pleak}. Since the downstream participants of LLM-based applications are users, these risks can cause direct harm in real-world scenarios.

In summary, risks are pervasive throughout the entire LLM supply chain, and understanding their dependencies and threats is critical for facilitating the security assurance of LLM systems.

\begin{figure*}[t]
    \centering
    \includegraphics[width=\textwidth]{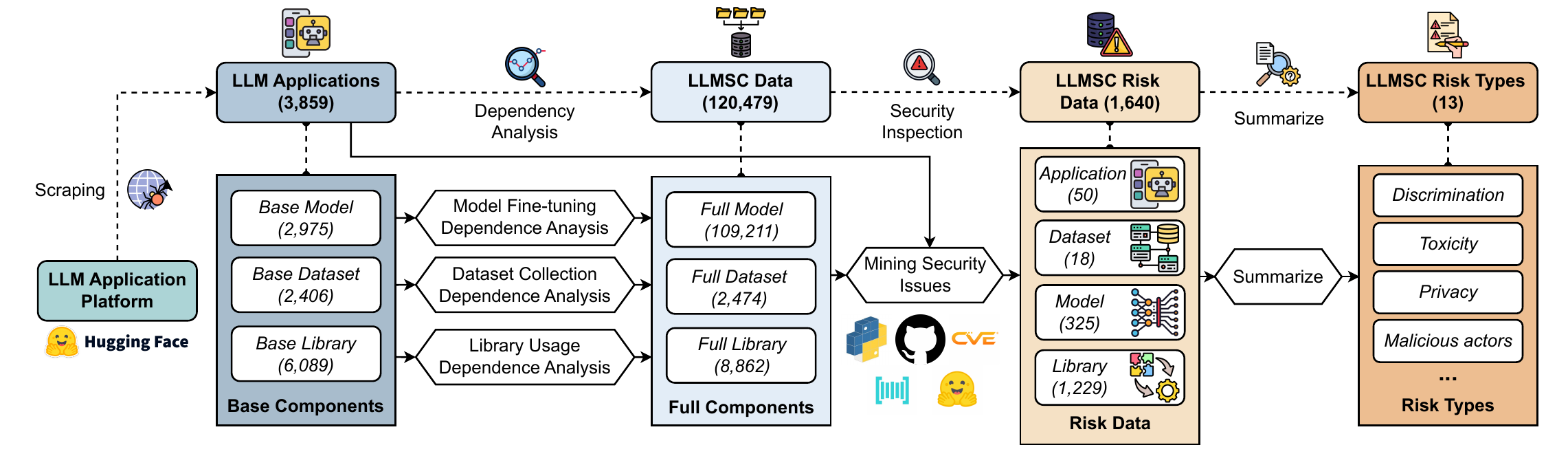}  
    \caption{The workflow of \dataset construction.}
    \label{fig:workflow}
\end{figure*}


\section{\dataset Construction}
\label{sec:dataset}

In this section, we introduce the construction process of our benchmark datasets, which is divided into two main parts: dependency dataset collection and risk issues extraction. The overview of our collection framework is shown in Figure~\ref{fig:workflow}.

\subsection{LLMSC Dependency Dataset Collection} 
The basic LLMSC dependency dataset collection mainly includes the application collection, the collection of base models, datasets and libraries used in applications, and their dependent components collection. Specifically, the detailed dataset construction process of \dataset is summarized as follows:

\textit{Step 1: LLM-enabled application collection.} We collect applications building upon LLMs from the famous LLM application database, Hugging Face \cite{huggingface-Spaces}. Hugging Face provides thousands of LLM applications uploaded by developers from different domains. We consider all these domains, which cover 36 categories based on their usage scenarios, such as \textit{StyleTransfer} and \textit{FineTuningTools}.

In total, 3,859 LLM-enabled applications are collected and included in \dataset, each of which provides 11 types of basic information~(shown in Table~\ref{tab:information_app}), including the creator, popularity, and the description of the application.

\begin{table}[h]
\centering
\caption{Basic information of LLM-enabled applications.}
\resizebox{0.8\textwidth}{!}{ 
\begin{tabular}{cc}
\hline
Information & Description \\ \hline
APP\_name & the name of the application \\
creator & the Hugging Face username of the application creator \\
url & the URL link to the application page \\
description & short description of the application \\
app\_text & full text from the application page~(e.g., README or UI content) \\
requirements & \begin{tabular}[c]{@{}c@{}}the \textit{requirements.txt} file included in the application project directory \\ Used to specify the Python libraries required to run the app\end{tabular} \\
directory & the project directory structure of the application \\
likes & number of likes received \\
community & number of discussions and the titles of all discussions \\
linked\_models & integrated large language models \\
commit & \begin{tabular}[c]{@{}c@{}}includes the number of commits, the time of the first commit,\\ and the time of the last commit~(up to the extraction date)\end{tabular} \\ \hline
\end{tabular}
}
\label{tab:information_app}
\end{table}

\textit{Step 2: Model collection.} Based on the collected LLM-based applications, we further extract both the base models and their dependent models. To identify base models, we primarily rely on the \textit{linked\_model} field, which indicates the models directly used by each application. Among the 3,859 applications, 2,008 explicitly provide \textit{linked\_model} information, from which we extract a total of 2,975 base models.

\begin{figure}[t]
	\centering
    \includegraphics[width=0.5\textwidth]{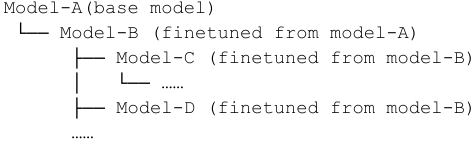}
    \caption{Template of model fine-tuning dependency chains. }
    \label{figure:model_dependency_template}
\end{figure}

To analyze the model-level supply chain, we construct model dependency chains based on fine-tuning relationships. We leverage the Hugging Face model pages, which typically include metadata indicating whether a model is fine-tuned from another, to identify transitive dependencies. Specifically, we develop an automated recursive crawler that traces fine-tuning relationships in both directions: (1) models fine-tuned from a given base model, and (2) models from which the base model was fine-tuned. Figure~\ref{figure:model_dependency_template} illustrates the structure of the resulting model fine-tuning dependency chains. 

In total, \dataset covers 109,211 distinct LLMs, comprising 2,975 base models and 106,236 dependent models, along with 2,275 fine-tuning dependency chains. \footnote{A fine-tuning dependency chain refers to a sequence of models where each model is fine-tuned from its immediate predecessor.}

\textit{Step 3: Dataset collection.} In LLM-based applications, datasets serve as a critical dependency, often used to train or fine-tune the underlying models. However, identifying dataset usage is challenging, as applications are not required to explicitly expose this information. Initially, we were able to collect dataset information for only 308 applications that directly mentioned their datasets.

To expand coverage, we further analyzed the descriptions of associated models. To support reliable extraction, we developed an LLM-assisted and human-confirmed mechanism. Specifically, given the application or model descriptions, we prompt an LLM to identify potential datasets used, and then manually verify the correctness of its responses. 
Using this approach, we successfully identified 2,406 datasets as base datasets. We then analyzed their origins and dependencies, identifying 68 dataset dependency chains that trace how certain base datasets are derived from others (i.e., a subset). These chains led to the identification of 68 dependent datasets.

In total, \dataset covers 2,474 distinct datasets, comprising 2,406 base datasets and 68 dependent datasets, along with 68 dataset dependency chains.

\textit{Step 4: Library collection.} As highlighted by prior studies~\cite{wang2025large,wang2025sok}, potential risks in LLM applications can often stem from the underlying implementation, such as vulnerabilities introduced through modified or misused APIs. To account for this, we analyze the libraries used in each application. We begin by extracting the base libraries from the requirements files provided by each LLM application, resulting in 6,089 version-specific libraries.

To capture the broader supply chain, we perform a recursive dependency analysis to identify dependent libraries, i.e., those indirectly included through transitive dependencies. Specifically, for each base library, we query and resolve its dependency relationships until reaching base packages with no further dependencies. This process allows us to construct a complete view of library calling dependency chains.

In total, \dataset includes 8,862 libraries, comprising 6,089 base libraries directly specified by applications and 2,773 dependent libraries, along with their corresponding 8,862 library calling dependency chains.

\begin{tcolorbox}[size=title,opacityfill=0.1]
\noindent
\textbf{Finding 1:} Through our data collection process, we found that library dependencies are relatively easy to identify from dependency files (e.g., requirements.txt). In contrast, identifying dependencies for models and datasets is more challenging, and our current analysis largely relies on user-provided descriptions. This limitation highlights the need for more advanced and automated dependency analysis techniques to capture the full scope of dependencies within LLM-based systems.

\end{tcolorbox}

\subsection{Risk Issue Extraction}

To support comprehensive risk investigation within the LLM supply chain, we extract potential risk issues associated with both the applications and each component of the LLMSC. For common software libraries, vulnerabilities can be readily identified using publicly available CVE databases such as OSV.dev. However, for user-developed applications and other components like models and datasets, there are no standardized public resources that systematically document risk information.

To address this gap, we follow the methodology of prior work~\cite{quan2022towards, yu2024bugs} by mining risk-related information from user discussions on platforms such as GitHub Issues. We then manually review and categorize the extracted issues into predefined risk types. The detailed processes for collecting and categorizing risk issues for applications, models, and datasets are introduced as follows:

\begin{table}[!t]
\centering
\caption{Categories of critical risks in LLMs}
\label{tab:keywords}
\resizebox{.9\textwidth}{!}{
\begin{tabular}{cc}
\hline
Risk & Explanation \\ \hline
discrimination/biasness & \begin{tabular}[c]{@{}c@{}}Bias or unfair treatment in model outputs or dataset \\ based on gender, race, or other attributes.\end{tabular} \\
toxicity & Offensive, harmful, or inappropriate language generated by the model. \\
privacy & \begin{tabular}[c]{@{}c@{}}Risks of exposing sensitive or personally identifiable information \\ in model responses or training data.\end{tabular} \\
security & \begin{tabular}[c]{@{}c@{}}General concerns related to vulnerabilities, system breaches, \\ or unauthorized access.\end{tabular} \\
misinformation & Factual inaccuracies in model outputs that may mislead users. \\
malicious actors & Entities that exploit the model for harmful or unauthorized purposes. \\
misuse & Improper or unintended use of the model, potentially causing harm. \\
hallucination & Confidently generated but factually incorrect or fabricated information. \\
catastrophic forgetting & \begin{tabular}[c]{@{}c@{}}Model forgetting previously learned information \\ after fine-tuning or continual learning.\end{tabular} \\
cybercrime & \begin{tabular}[c]{@{}c@{}}Use of the model to assist in illegal online activities \\ such as phishing or fraud.\end{tabular} \\
trustworthiness & \begin{tabular}[c]{@{}c@{}}Degree to which users can rely on the model’s outputs\\  to be accurate and safe.\end{tabular} \\
adversarial attack & \begin{tabular}[c]{@{}c@{}}Inputs crafted to fool the model into producing\\  incorrect or unintended outputs.\end{tabular} \\
prompt hacking & \begin{tabular}[c]{@{}c@{}}Manipulation of prompts to trick the model into behaving \\ in unexpected or harmful ways.\end{tabular} \\ \hline
\end{tabular}
}
\end{table}

\textit{Step 1: Risk type identification.}  Following prior works~\cite{slattery2024ai, das2025security}, we first identify 13 common-used types of risks that developers should be aware of, as shown in Table~\ref{tab:keywords}. These risk categories serve as the foundation for subsequent issue extraction and classification.

\textit{Step 2: Risk issue extraction.} Using the identified risk types as keywords, we extract potential risk-related information for different components. For applications and models, we gather data from two main sources: Hugging Face Discussions and GitHub Issues. For datasets, we limit our extraction to Hugging Face Discussions, as most datasets do not have dedicated GitHub repositories. This process yields an initial collection of 459, 3,919, and 29 risk issue candidates for applications, models, and datasets, respectively.

\textit{Step 3: Risk issue cleaning.} Since keyword-based extraction can result in noisy or irrelevant entries, we perform a manual verification step to ensure the quality of the identified risk issues. Three authors independently review the extracted items, and disagreements are resolved through discussion until consensus is reached. After this cleaning process, we retain 50, 325, and 18 validated risk issues for applications, models, and datasets, respectively.

For the risk information extraction of libraries, we perform vulnerability matching on all extracted dependency packages and their version information in multiple mainstream vulnerability databases~(CVE, PyPI Security Notice, OSV.dev, GitHub Advisory Database). By comparing dependencies with disclosed security vulnerability records, we identify a total of 1,229 CVE vulnerability records related to these libraries.

\begin{tcolorbox}[size=title,opacityfill=0.1]
\noindent
\textbf{Finding 2:} Although it is easy to find GitHub issues and Hugging Face discussions that include risk-relevant comments, only a few of them~(8.9\%) can be treated as real-world risk issues after manual checking. there is a pressing need for a dedicated and structured risk database for models and datasets, similar to the CVE system.
\end{tcolorbox}

\textbf{Dataset Comparison.} We further compare \dataset{} with existing datasets in terms of their coverage of applications, components, and associated risk information, as summarized in Table~\ref{tab:dataset}. \dataset is the first benchmark that comprehensively captures multiple components of the LLM supply chain while also documenting their interdependencies and associated risk issues.

\begin{table}[!t]
\centering
\caption{Comparison of LLMSC Datasets. }
\label{tab:dataset}
\resizebox{0.8\textwidth}{!}{
\begin{tabular}{c|ccccc}
\hline
\textbf{Benchmark} & \textbf{Application} & \textbf{Model} & \textbf{Dataset} & \textbf{Library} & \textbf{Risk Issues}  \\ \hline
SoK~\cite{wang2025sok} & $\checkmark$ & $\times$ & $\times$ & $\times$ & $\checkmark$ \\ \hline
Hydrangea~\cite{shao2024llms} & $\checkmark$ & $\times$ & $\times$ & $\times$ & $\checkmark$ \\ \hline
GPTZoo~\cite{hou2024gptzoo} & $\checkmark$ & $\times$ & $\times$ & $\times$ & $\times$ \\ \hline
Hou \emph{et al.}~\cite{hou2025security} & $\checkmark$ & $\times$ & $\times$ & $\times$ & $\checkmark$ \\ \hline
Xie \emph{et al.}~\cite{xie2024llm} & $\checkmark$ & $\times$ & $\times$ & $\times$ & $\checkmark$\\ \hline
Su \emph{et al.}~\cite{su2024gpt} & $\checkmark$ & $\times$ & $\times$ & $\times$ & $\checkmark$  \\ \hline
BeeTrove~\cite{mafei2024beetrove} & $\checkmark$ & $\times$ & $\times$ & $\times$ & $\checkmark$  \\ \hline
Epic~\cite{epic2024gptstore} & $\checkmark$ & $\times$ & $\times$ & $\times$ & $\checkmark$  \\ \hline
\dataset~(Ours) & $\checkmark$ & $\checkmark$ & $\checkmark$ & $\checkmark$ & $\checkmark$  \\ \hline
\end{tabular}
}
\end{table}

\section{\dataset Analysis}
Based on \dataset, we conduct an empirical study to better understand the structure and risks of the LLM supply chain. Given that LLM-based applications represent a relatively new and rapidly evolving class of software systems, we focus our analysis on two key aspects: (1) the characteristics of the applications and their constituent components, and (2) the potential risks associated with each component. This two-fold analysis provides valuable insights into the complexity, dependency patterns, and security implications of modern LLM-based systems.

\subsection{Application Analysis}
\label{sec：application_analysis}

\begin{figure}[h]
	\centering
	\includegraphics[width=0.6\textwidth]{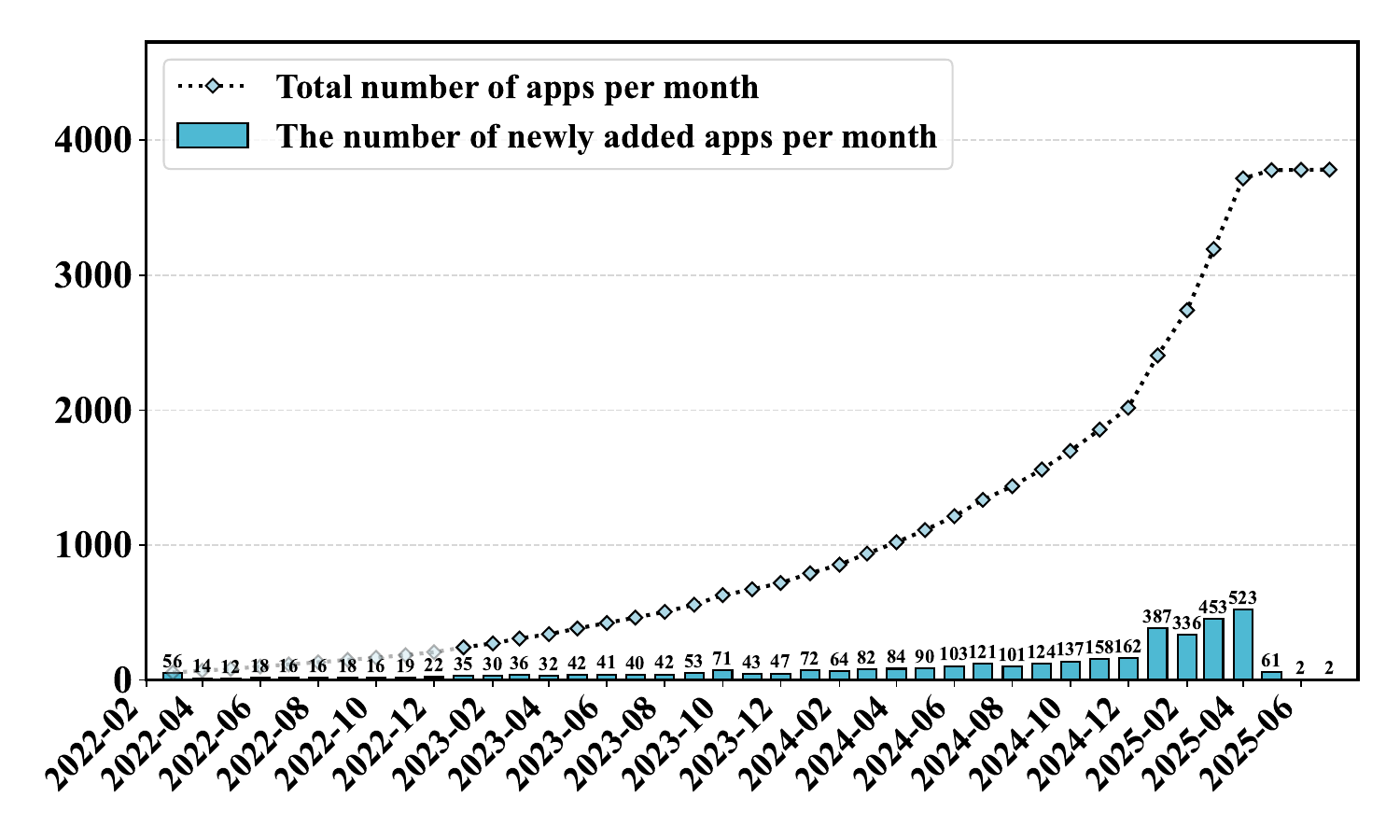}
	\caption{Number of models over months in \dataset~(years: 2022–2025).}
	\label{fig:trend_of_applications}
\end{figure}

\textbf{Characteristics Analysis.} Overall, the distribution of applications across categories is relatively balanced, with most categories containing approximately 130 applications. However, certain categories, such as StyleTransfer, FineTuningTools, and CharacterAnimation, contain fewer than 50 applications each. This observation suggests that while LLMs are being rapidly adopted across a wide range of tasks, some specialized domains are still in the early stages of development and may require further maturity in tooling, model support, or community engagement.

\begin{figure}[h]
	\centering
	\includegraphics[width=0.6\textwidth]{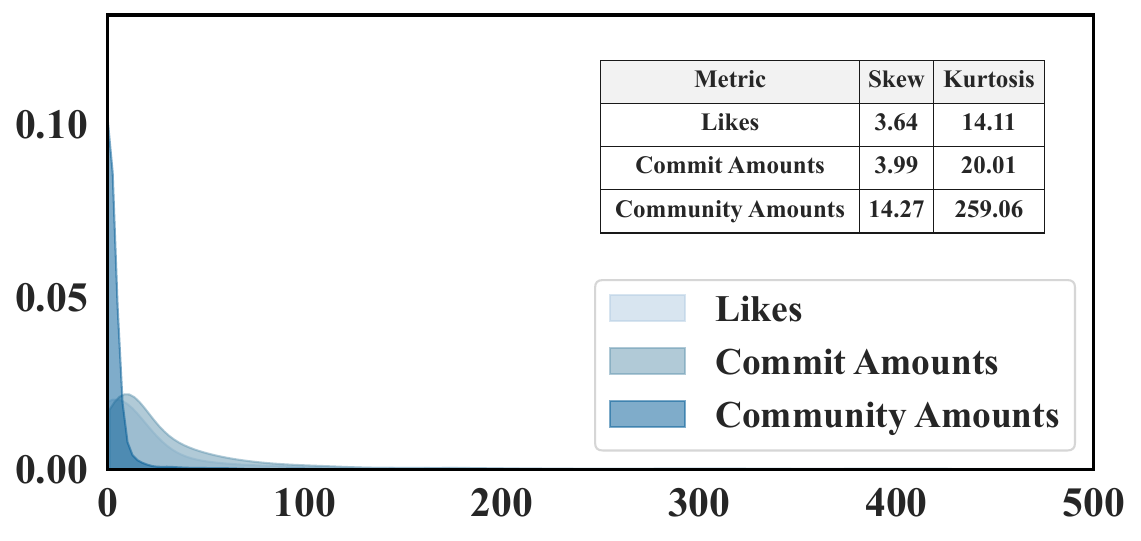}
	\caption{Density plots of app likes, commit counts, and community activity.}
	\label{fig:popularity_application}
\end{figure}

\textit{Development trend of LLM-enabled applications.} Figure~\ref{fig:trend_of_applications} shows the development trend of our collected applications over months. We can see that there is a clear increasing trend for the LLM-enabled application construction. Only in 2025, 1,764 new applications have been released, accounting for 45.7\% of all applications, indicating the urgent need to establish a proper dataset to support the exploration of LLMSC. 

\begin{figure}[H]
	\centering
	\includegraphics[width=0.5\textwidth]{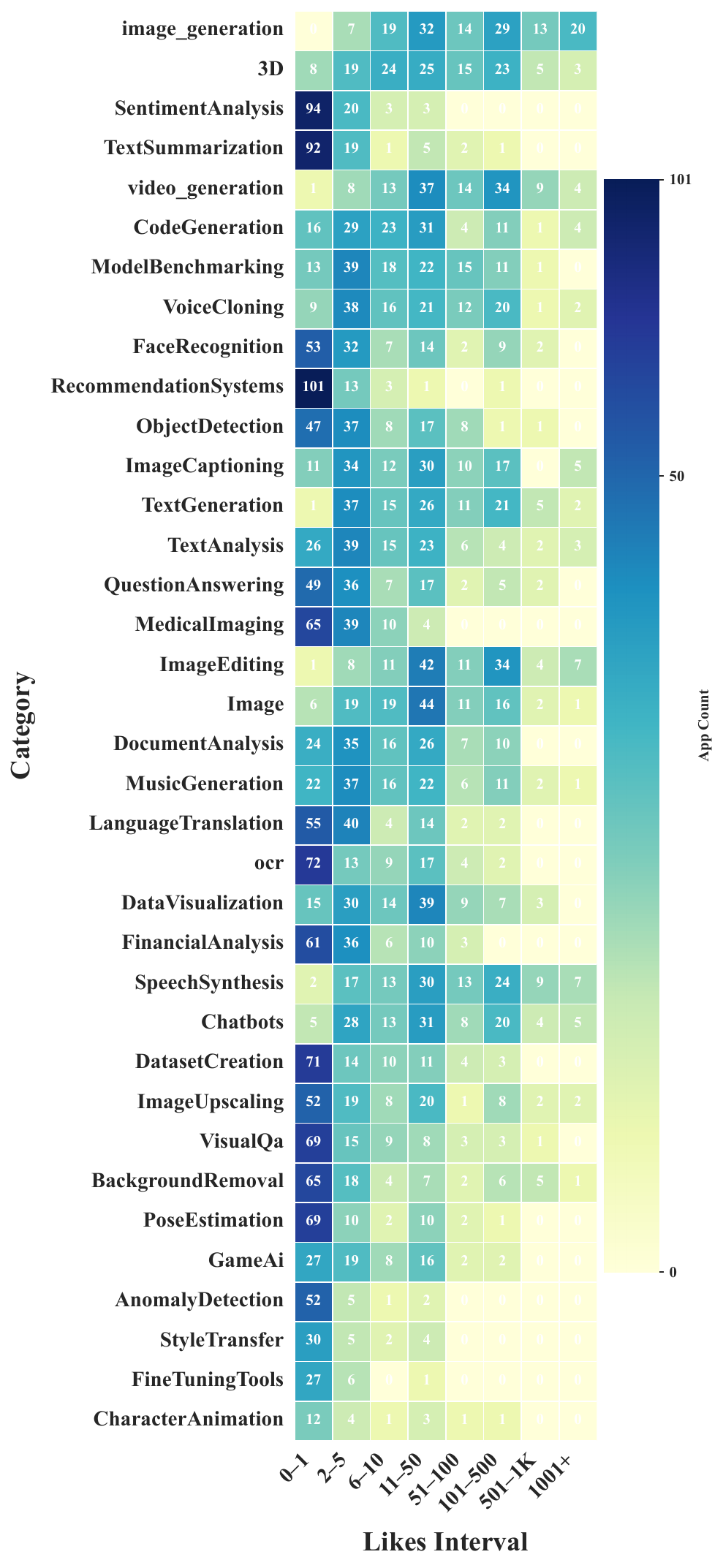}
	\caption{Distribution of likes considering different categories.}
	\label{fig:popularity_application_per_category}
\end{figure}

\textit{Popularity of applications.} Figure~\ref{fig:popularity_application} depicts the quantification~(by counting the number of likes, commits, and community discussions) of popularity of the collected applications. It indicates that although thousands of applications have been produced, most of them have not received much attention. For instance, 2,498~(63.6\%) applications obtained no more than 50 likes, only 64~(1.7\%) applications are recommended by more than 1000 users. We go deeper to analyze how the number of commits and the number of community discussions affect the popularity of applications by computing their correlation.  Correlation analysis based on the Pearson correlation coefficient indicates a weak positive correlation between the two numbers to the popularity of applications with significance~(P-value < 0.01). This means more frequent application updates and more discussions lead to popular applications.

In addition, we investigate the popularity of each category of applications. The results~(shown in Figure~\ref{fig:popularity_application_per_category}) demonstrate that applications in the \textit{image\_generation} category generally have a high number of likes and are currently the most popular application type on Hugging Face. We conjecture that this phenomenon is due to the popularity of AI painting~(such as Diffusion models). In addition, \textit{SpeechSynthesis}, \textit{ImageEditing}, \textit{video\_generation}, \textit{TextAnalysis}, and \textit{Chatbots} are also relatively more popular. Most natural language processing related applications, such as \textit{RecommendationSystems}, \textit{SentimentAnalysis}, and \textit{TextCummarization} receive less attention.


\begin{figure}[h]
    \includegraphics[width=0.9\textwidth]{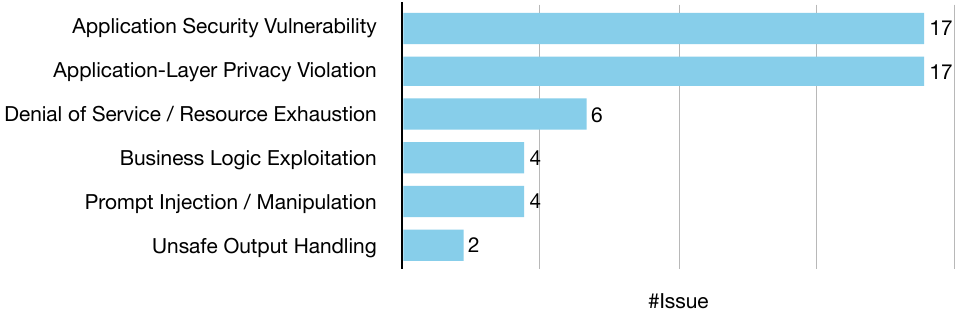}
	\caption{Collected risk issues in applications}
	\label{fig:application_security_issue}
\end{figure}

\begin{tcolorbox}[size=title,opacityfill=0.1]
\noindent
\textbf{Finding 3:} There is a clear upward trend in the development of LLM-enabled applications, with image generation-related applications receiving notably higher user engagement. This also highlights the growing importance of performing comprehensive risk analysis on these emerging software systems to ensure their safety and trustworthiness.
\end{tcolorbox}

\textbf{Risk Analysis.} By our risk extraction and manual analysis on the application-specific risk data, we identify 50 issues spanning six categories. 

\begin{itemize}[leftmargin=*]
    \item \textit{Application Security Vulnerability: }Classic web or application security issues within LLM applications, such as insecure API endpoints, cross-site scripting, or improper storage of user data.
    \item \textit{Application-Layer Privacy Violation: }Privacy breaches caused by the application itself, such as exposing other users' data, leaking personally identifiable information, or mishandling data shared with third-party services.
    \item \textit{Denial of Service / Resource Exhaustion: }A user can intentionally craft inputs that cause the application to consume excessive resources~(e.g., API calls), affecting availability for other users.
    \item \textit{Business Logic Exploitation: }The LLM is manipulated to abuse or circumvent the intended business rules of the application, such as tricking a system into issuing unauthorized actions.
    \item \textit{Prompt Injection / Manipulation: }Vulnerabilities where malicious user input hijacks the core logic of the LLM application, causing it to ignore instructions, leak system prompts, or perform unauthorized actions.
    \item \textit{Unsafe Output Handling: }Cases where the application unsafely trusts LLM outputs, leading to vulnerabilities, injection attacks, or flawed decisions.
\end{itemize}

\begin{figure}[h]
	\centering
	\includegraphics[width=0.8\textwidth]{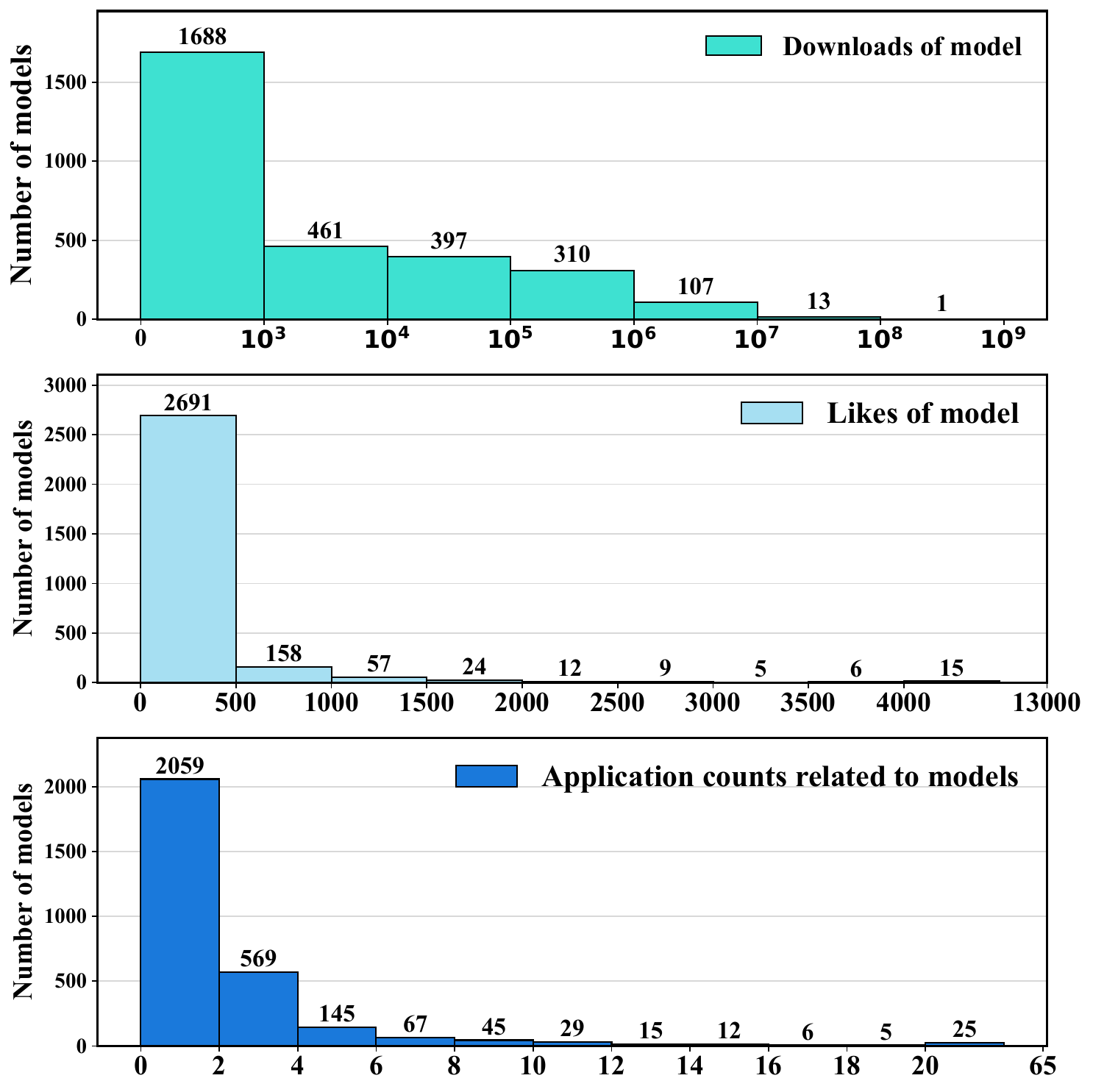}
	\caption{Distribution of model popularity}
	\label{figure:model_popularity}
\end{figure}
As shown in Figure~\ref{fig:application_security_issue}, the results demonstrate that most of the risks~(73.84\%) lie in \textit{Application Security Vulnerability} and \textit{Application-Layer Privacy Violation} two categories, highlighting the necessity of more strict security and privacy assessment of LLM-enabled applications. On the other side, four risks come from the \textit{Prompt Injection / Manipulation} category, which are manually introduced by attackers, indicating the need for stronger defense techniques for the Hugging Face platform. 

\textit{Case study.} We showcase a representative risk case~\cite{chatui299} in the category of \textit{Application-Layer Privacy Violation} in detail. Specifically, in the application \textit{huggingchat/chat-ui}, a newly registered user~(User A) creates a Hugging Face account and initiates their first conversation with the Falcon model. Without any prior interactions, User A unexpectedly observes four chat histories from other users in their account. This case highlights a potential privacy vulnerability where chat histories from other users were inadvertently exposed, or testing data was not properly cleared, resulting in unintended data leakage.

\begin{tcolorbox}[size=title,opacityfill=0.1]
\noindent
\textbf{Finding 4:} As revealed by our analysis, \textit{Application Security Vulnerability} and \textit{Application-Layer Privacy Violation} are two representative risks with more than 73.84\% reported issues among all issues.
\end{tcolorbox}

\subsection{Model Analysis}

We further analyze 2,975 models that are directly used by our collected 3,859 applications.

\textbf{Characteristics Analysis.} Figure ~\ref{figure:model_popularity} illustrates the number of applications that utilize each model, the number of likes, and the number of downloads. Among these models, 2059 models are used by a single application. Some models, are relatively more frequently integrated into applications, reflecting their general-purpose capability and ecosystem influence. For example, \textit{black-forest-labs/FLUX.1-dev} is used by 64 applications, and \textit{openai/clip-vit-large-patch14} is used by 60 applications. By a correlation analysis where the results, we found positive correlations between model utilization times, the number of model downloads, and the number of likes, these three factors, with significance~(P-value < 0.01). 

\begin{figure}[h]
	\centering
	\includegraphics[width=0.6\textwidth]{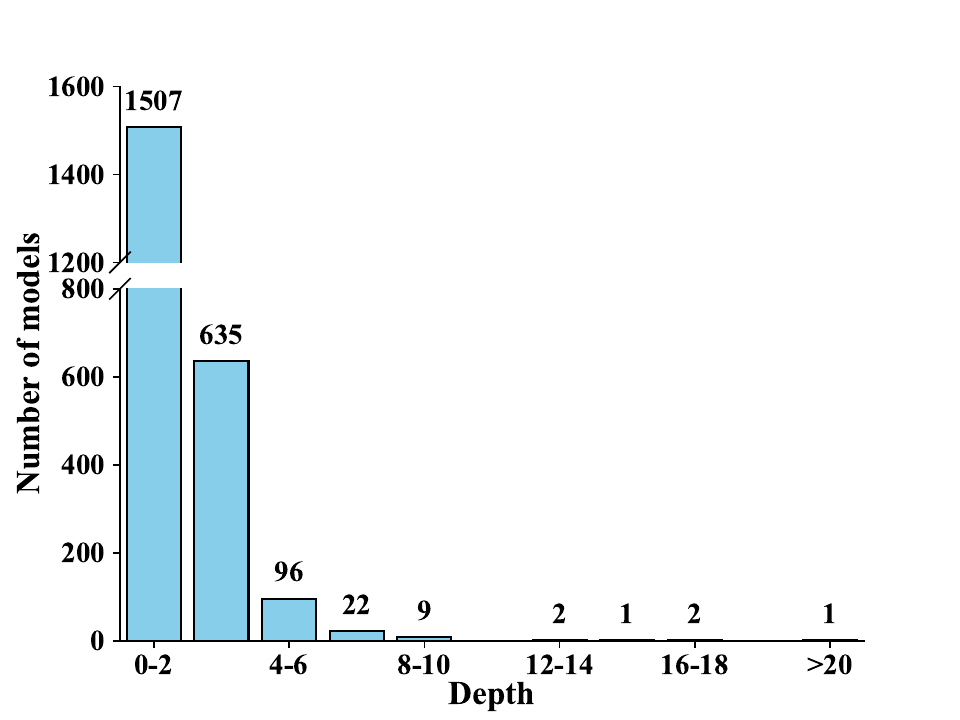}
	\caption{Distribution of the depth of model  dependency}
	\label{fig:model_dependency_depth}
\end{figure}

\textit{Distribution of model dependency graphs.} Based on the construction of the model fine-tuning chain, we collect a total of 109,211 models that are in the LLMSC. Figure~\ref{fig:model_dependency_depth} shows the distribution of these dependency chains. The results demonstrate that most graphs are at the depth levels of 0, 1, and 2, indicating that most models are either standalone~(i.e., not further fine-tuned) or fine-tuned only for one or two generations. Only a small number of models form deep fine-tuning chains. For example, the model \textit{CohereLabs/c4ai-command-r-v01} reaches a depth of 40, and both \textit{openai-community/gpt2} and \textit{google-t5/t5-small} have dependency depths greater than 15.

Regarding dependency width, 26 root models have subtree widths exceeding 1,000, indicating that these models are extensively reused within the Hugging Face ecosystem. Additionally, 161 base models have subtree widths greater than 100, forming long chains of fine-tuned descendants. Collectively, these 161 base models give rise to 98,436 models, representing 90. 1\% of all models analyzed. These results demonstrate that although the total number of models on Hugging Face is large, they can be traced back to a relatively small set of foundational root models. This phenomenon is highly consistent with observed trends in model reuse across the natural language processing community~\cite{bommasani2021opportunities}, underscoring the central role of foundation models such as LLaMA, BERT, and Mistral in shaping the model ecosystem.

\begin{tcolorbox}[size=title,opacityfill=0.1]
\noindent
\textbf{Finding 5:} Most of the base models in our dataset (42.8\%) are not root models—that is, they are fine-tuned from other models. Notably, the maximum depth of fine-tuning chains reaches up to 40, illustrating deeply nested dependencies. Furthermore, 98,436 models (90.1\%) can be traced back to just 161 core root models, highlighting the central role of a small number of foundation models in the LLM ecosystem. These findings demonstrate the high degree of model reuse and interdependence, underscoring the critical importance of supply chain risk analysis.

\end{tcolorbox}

\begin{figure}[h]
    \centering
    \includegraphics[width=0.8\textwidth]{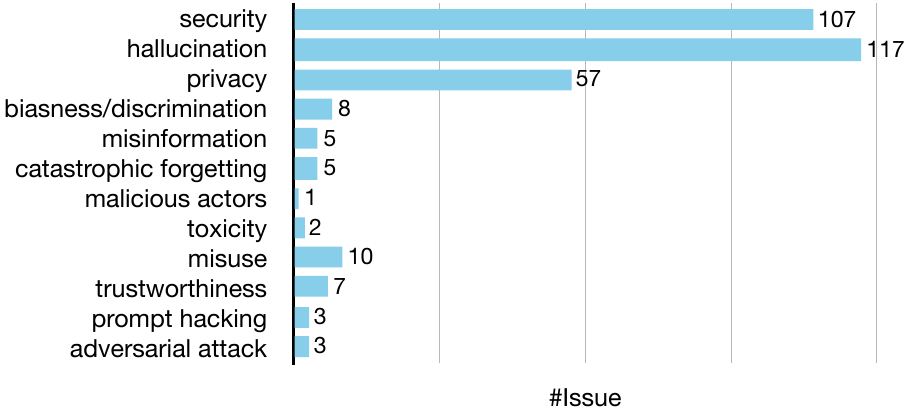}
	\caption{Collected risk issues in models}
	\label{fig:model_security_issue}
\end{figure}

\textbf{Risk Analysis.} After the risk issue extraction, we filter out 325 relevant issues across 12 risk categories~(the description of these categories can be found in Table~\ref{tab:keywords}) revealed by the model users. Figure~\ref{fig:model_security_issue} presents the distribution of these issues. We can see that \textit{security}, \textit{hallucination}, and \textit{privacy} are the three most common issues, accounting for 86.46\% of total issues. Interestingly, adversarial attack, which is a hot topic in deep learning security, only counts for 3 issues. 

\begin{figure}[h]
\centering
\includegraphics[width=0.8\textwidth]{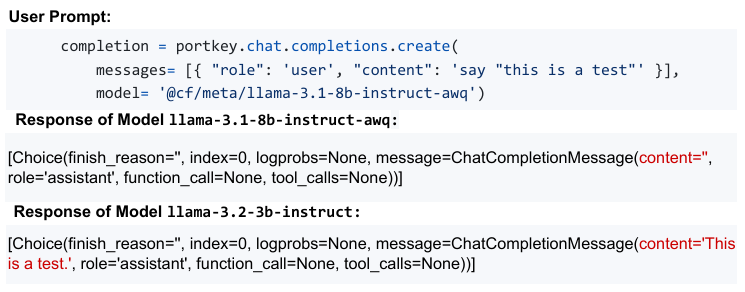}
	\caption{A case of LLM hallucination risk}
\label{fig:model_issue_example}
\end{figure}

\textit{Case study.} Figure~\ref{fig:model_issue_example} illustrates a case of the hallucination issue. For the same user prompt, \textit{llama-3.1-8b-instruct-awq} sometimes responds normally, but sometimes gives no response at all. In contrast, the \textit{llama-3.2-3b-instruct} model consistently works well and responds to the instruction with ``this is a test'' as expected.

\begin{tcolorbox}[size=title,opacityfill=0.1]
\noindent
\textbf{Finding 6:} Similar to the risk distribution observed in applications, \textit{security} is the most prevalent risk found in models. The second most common risk is \textit{hallucination}, indicating a strong need for effective model output validation and post-processing. Given the diverse range of risks associated with models, it is crucial to analyze not only their direct impact on the applications in which they are used, but also their potential to propagate downstream—affecting other models and applications that depend on them. 
\end{tcolorbox}

\subsection{Dataset Analysis}

In addition to the models, we also analyze the datasets used by the applications. In total, \dataset covers 2,406 datasets.

\begin{figure}[h]
\centering
\includegraphics[width=0.8\textwidth]{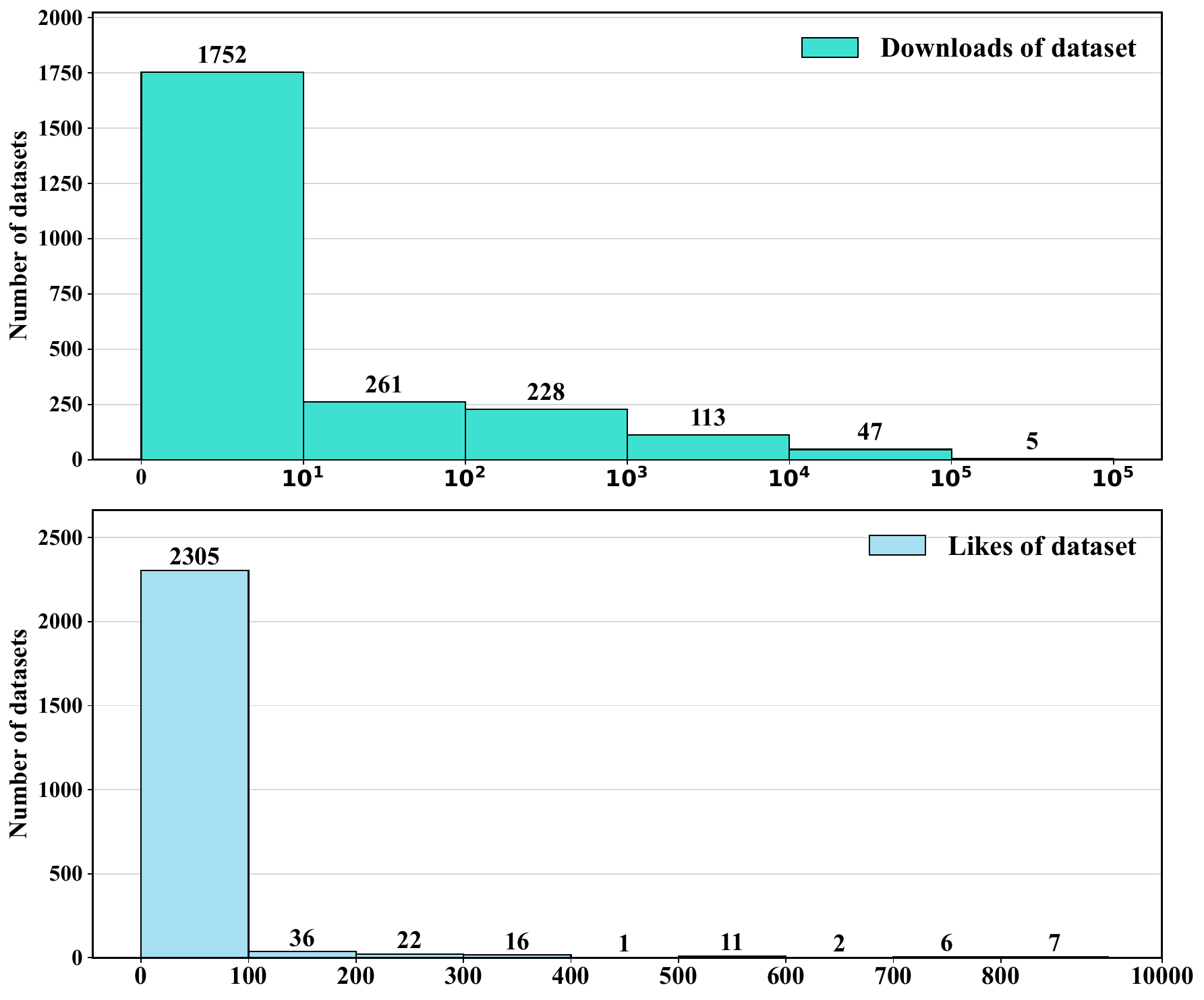}
	\caption{Distribution of dataset popularity}
\label{fig:dataset_distribution}
\end{figure}

\textbf{Characteristics Analysis.} Figure~\ref{fig:dataset_distribution} illustrates the distribution of downloads and likes of our collected datasets. Similar to the models, most datasets are downloaded rarely, e.g., 1752~(72.8\%) datasets are downloaded no more than 10 times. Some famous datasets, such as 
\textit{Salesforce/GiftEvalPretrain} and \textit{mlfoundations/dclm-baseline-1.0} gained huge attention and have been downloaded more than 200,000 times. We explore the relationship between download times and the number of likes and found a significant positive correlation between these two factors~(with P-value less than 0.01). 

\begin{figure}[h]
\centering
\includegraphics[width=0.6\textwidth]{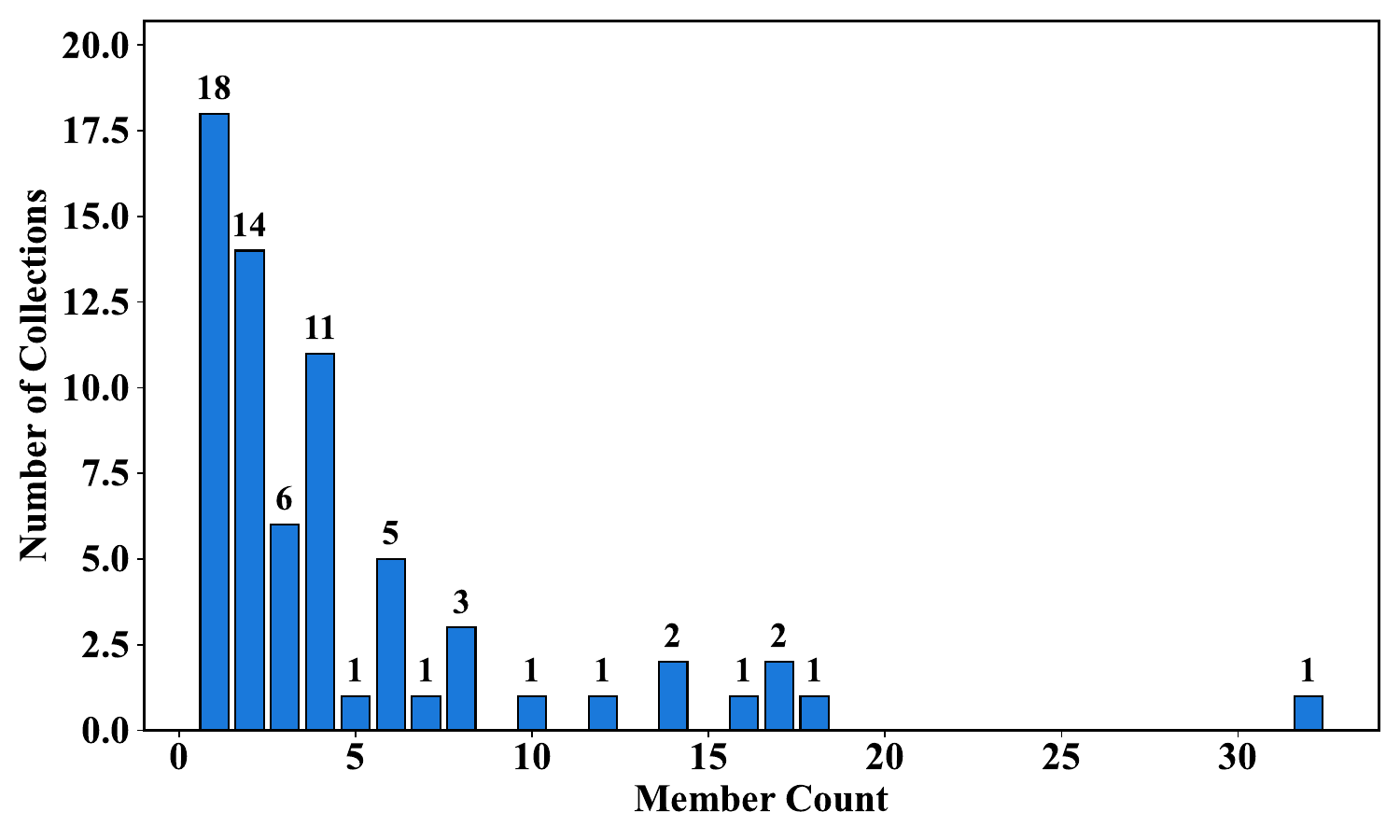}
\caption{Frequency distribution of the number of members.}
\label{fig:dataset_dependency}
\end{figure}

\textit{Distribution of dataset dependency graphs.} Among the 2,406 datasets, only 68 have the upstream~(or downstream) dataset collections. The depth of all data collection dependency graphs is 2, indicating that no datasets evolve multiple times. Figure~\ref{fig:dataset_dependency} shows the distribution of this collection of dependency graphs. The results follow a similar trend to the models and applications that most collections have single-child datasets. Only one collection, \textit{Tulu 3 Datasets}, is different from the others, and contains 32 datasets.

\begin{tcolorbox}[size=title,opacityfill=0.1]
\noindent
\textbf{Finding 7:} In contrast to models, the dependency information for datasets is considerably more obscure. We are only able to identify 68 dataset dependency relationships, accounting for just 2.82\% of all datasets. This limited visibility is primarily due to the lack of clear documentation or standardized recording of dataset usage statistics. The absence of such information highlights a significant gap in current practices and underscores the need for improved dataset dependency management to enable comprehensive supply chain assessment.
\end{tcolorbox}

\textbf{Risk analysis.} 
After the analysis, the dataset-related risk issues can be categorized as:

\begin{itemize}[leftmargin=*]
    \item \textit{Memory Exhaustion: }Excessive memory consumption during dataset loading.
    \item \textit{Cross-Platform Compatibility: }Inconsistent dataset loading across different platforms.
    \item \textit{Dataset Access: }Datasets cannot be accessed due to vulnerabilities in the Hugging Face Datasets Library or unavailable data sources.
    \item \textit{Incorrect Result: }The dataset can be loaded, but yields incorrect results.
    \item \textit{Parsing Error: }The dataset cannot be correctly parsed due to data format errors or loading script errors.
    \item \textit{Annotation Error: }The dataset contains mislabeled data.
    \item \textit{Data Noise: }The dataset contains many entries or strings that are irrelevant to the task.
    \item \textit{Includes Potential Sensitive Information: }The dataset contains users' sensitive information.
    \item \textit{Missing Data: }The dataset suffers from missing critical and widely-used entries.
\end{itemize}

We identified a total of 18 dataset-related issue cases, which can be categorized into two high-level types: dataset loading errors and dataset quality issues. Dataset loading errors include 3 cases of \textit{Dataset Inaccessibility}, 3 cases of \textit{Parsing Errors}, 2 cases of \textit{Cross-Platform Compatibility}, 1 case of \textit{Memory Exhaustion}, and 1 case of \textit{Incorrect Results}. Dataset quality issues comprise 4 instances of \textit{Data Noise}, 2 instances of \textit{Annotation Errors}, 1 instance of \textit{Potentially Sensitive Information}, and 1 case of \textit{Missing Critical Data}.

Data loading issues are typically caused by vulnerabilities in the Hugging Face datasets library. Dataset quality issues, on the other hand, usually stem from flaws in the data collection pipeline, which can lead to serious consequences, such as poisoning attacks against open-source LLMs~\cite{deepseek-issue}, making this a critical concern. These findings highlight the need for further improvement in the current dataset management mechanisms within Hugging Face.

\begin{tcolorbox}[size=title,opacityfill=0.1]
\noindent
\textbf{Finding 8:} We identify only a limited number of dataset-related risks, largely due to the lack of systematic management and reporting of dataset-specific vulnerabilities. However, the risks that are present reveal important insights—many of them, such as Missing Data, Sensitive Information Leakage, Data Noise, and Annotation Errors, have the potential to propagate through the LLM supply chain. This highlights the importance of dataset quality control and establishing mechanisms for tracking and mitigating dataset-originated risks.
\end{tcolorbox}

\subsection{Library Analysis}

\begin{figure}[h]
    \centering
    \begin{subfigure}[b]{0.5\textwidth}
        \includegraphics[width=\linewidth]{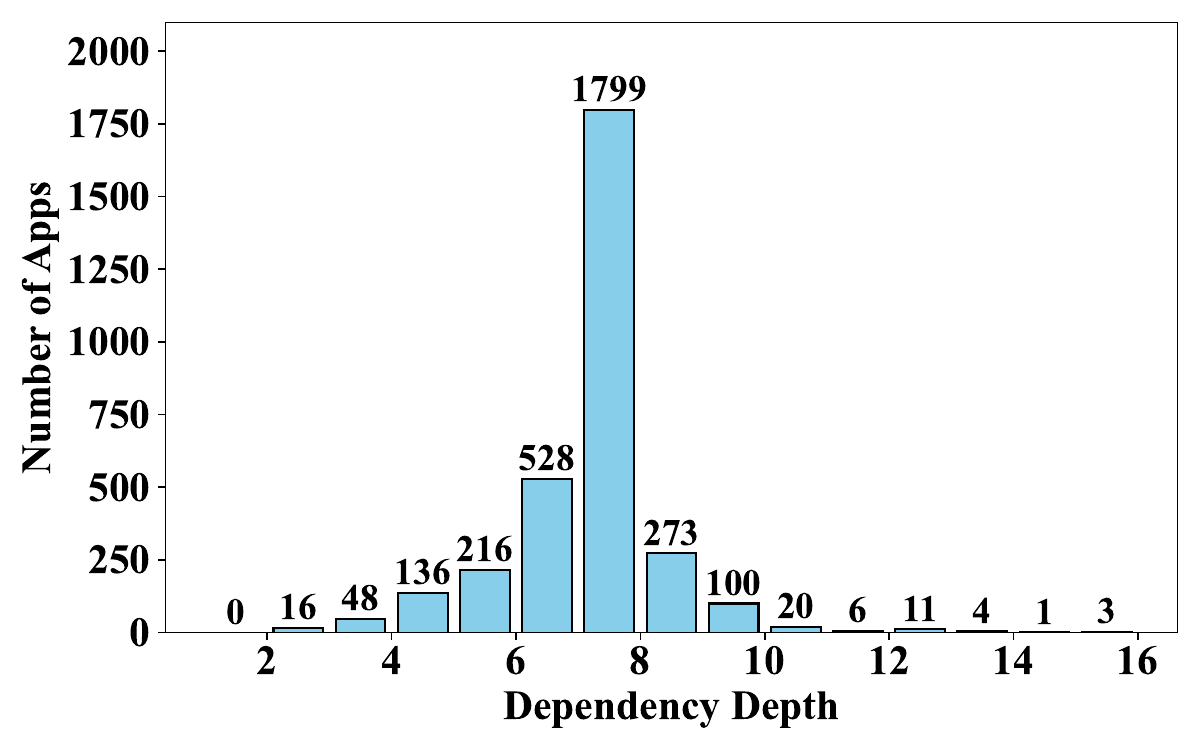}
        \caption{Library dependency depth.}
        \label{fig:library_dependency_depth}
    \end{subfigure}
    \hfill
    \begin{subfigure}[b]{0.5\textwidth}
        \includegraphics[width=\linewidth]{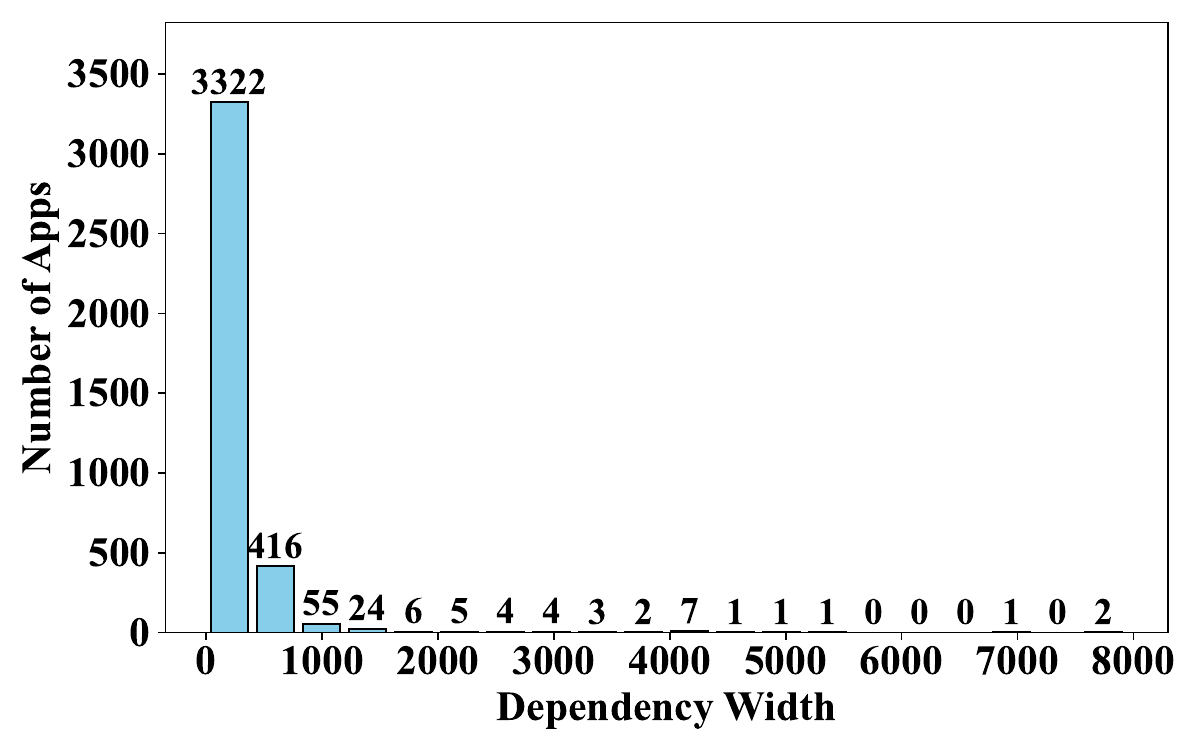}
        \caption{Library dependency width.}
        \label{fig:library_dependency_width}
    \end{subfigure}
    \caption{Library dependency distribution.}
    \label{fig:library_distribution}
\end{figure}

\textbf{Characteristic Analysis.} Figure~\ref{fig:library_distribution} shows the distribution of dependency depth and width. Most~(46.61\%) library dependency chains have a depth of 7, and the maximum depth is 15. Considering the applications, 693~(17.96\%) of them have a dependency width of 1, and application \textit{Tharindu1527/Stock-Market-Multi-Agent-System} has the maximum dependency width of 7943.

\begin{figure}[h]
\centering
\includegraphics[width=0.6\textwidth]{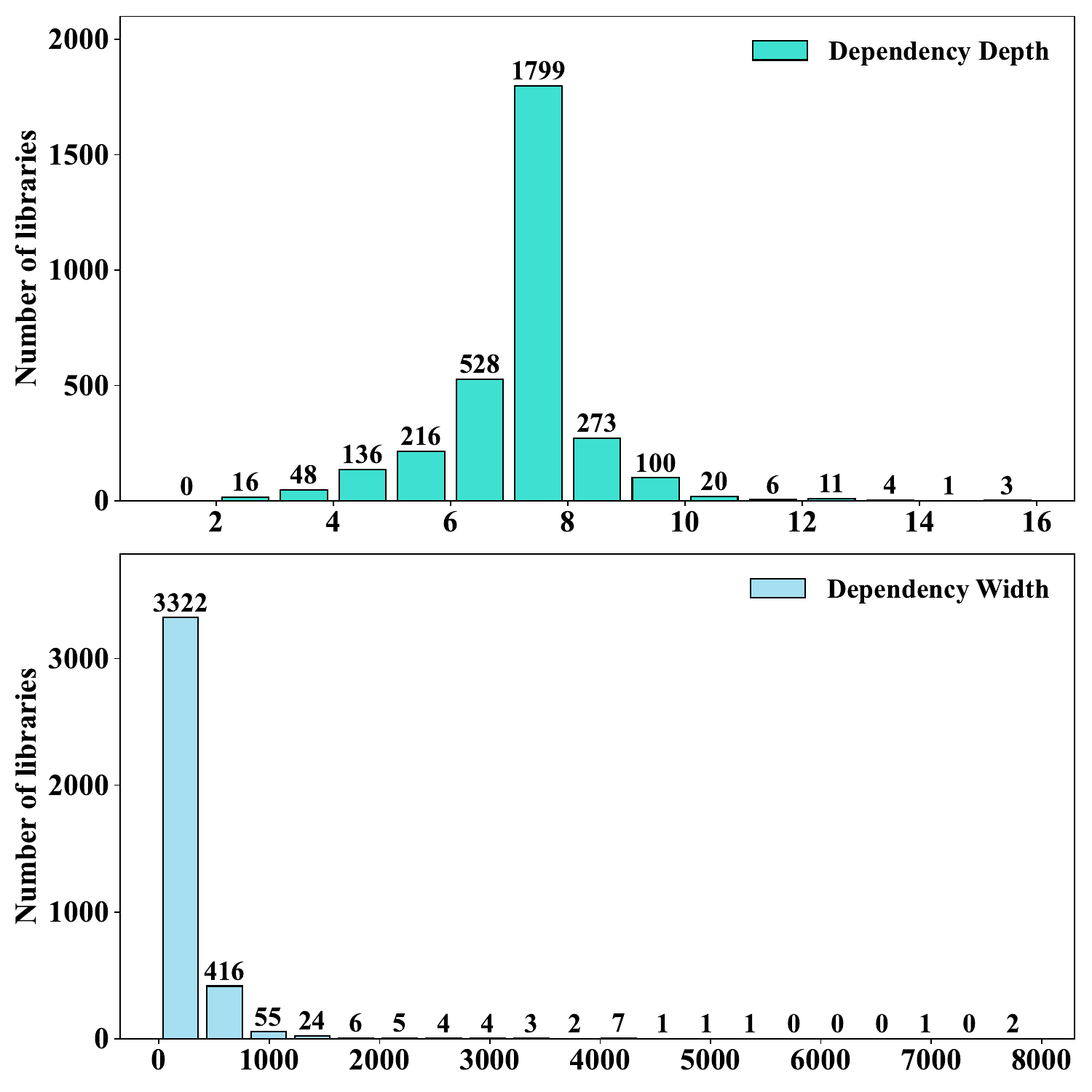}
\caption{Library dependency distribution.}
\label{fig:dataset_dependency_combined}
\end{figure}

\begin{figure}[h]
    \centering
    \includegraphics[width=0.8\textwidth]{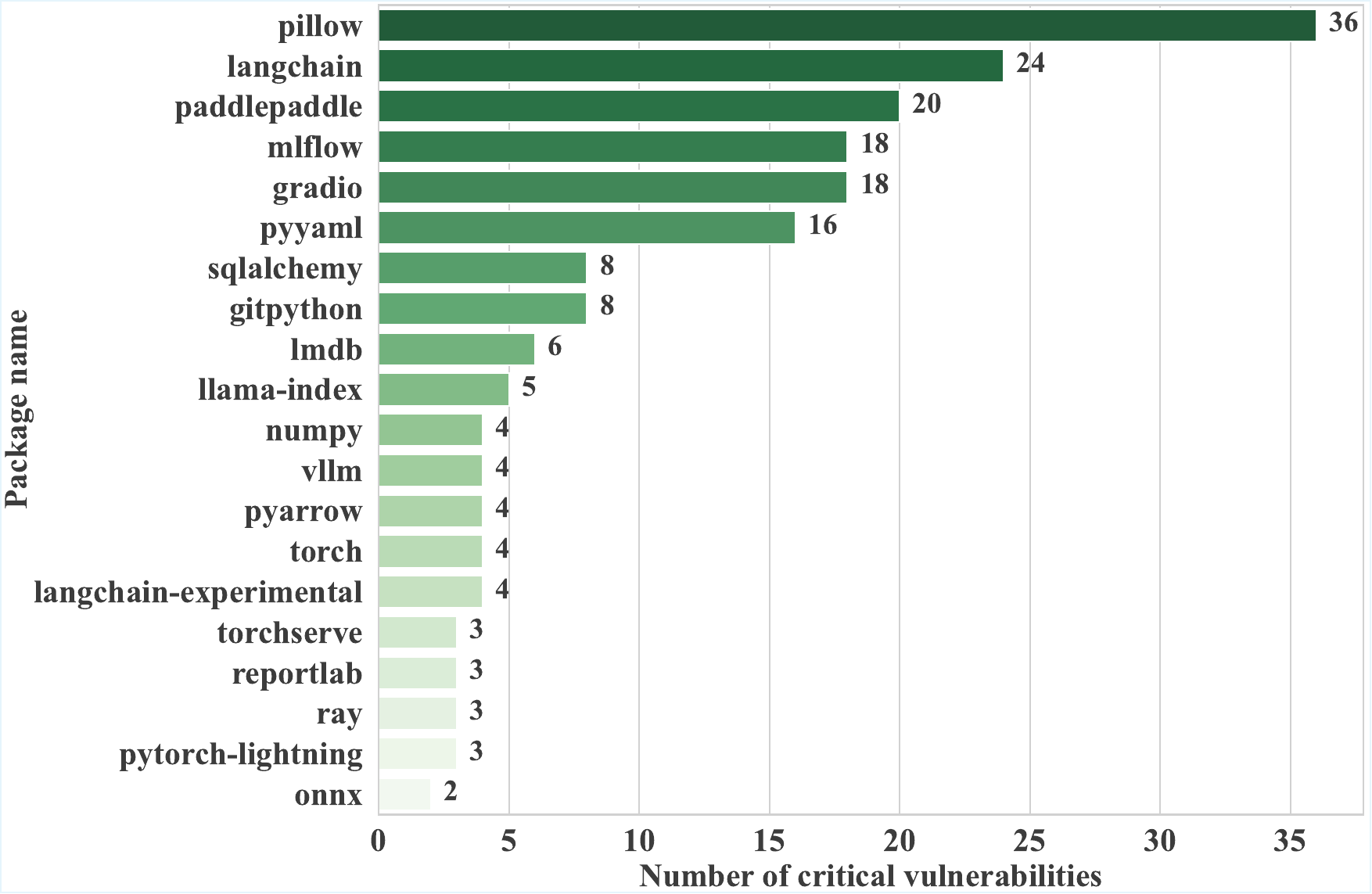}
	\caption{Top 20 libraries by critical CVE counts~(CVSS > 9).}
	\label{fig:critical_CVE_count}
\end{figure}

\textbf{Risk analysis.} First, we analyze the severity of libraries based on the CVSS scores. We found that \dataset contains 29 low-severity~($0 < \text{CVSS} < 4$), 365 medium-severity~($4 \le \text{CVSS} < 7$) , 543 high-severity~($7 \le \text{CVSS} < 9$), and 207 critical-severity~($9 \le \text{CVSS} \le 10$) CVEs. We go deeper to explore the libraries that contain critical vulnerabilities~($\text{CVSS} \ge 9$). The results shown in Figure~\ref{fig:critical_CVE_count} reveal that several key dependencies pose significant security risks. Notably, \textit{pillow}, \textit{langchain}, and \textit{paddlepaddle} have 36, 24, and 20 critical vulnerabilities, respectively, ranking at the top. Furthermore, among the 1,229 unique vulnerabilities, 77 have no available fixed version, while 1,152 have been addressed with a fix.

\begin{figure}[h]
    \centering
    \begin{subfigure}[b]{0.45\columnwidth}
        \includegraphics[width=\linewidth]{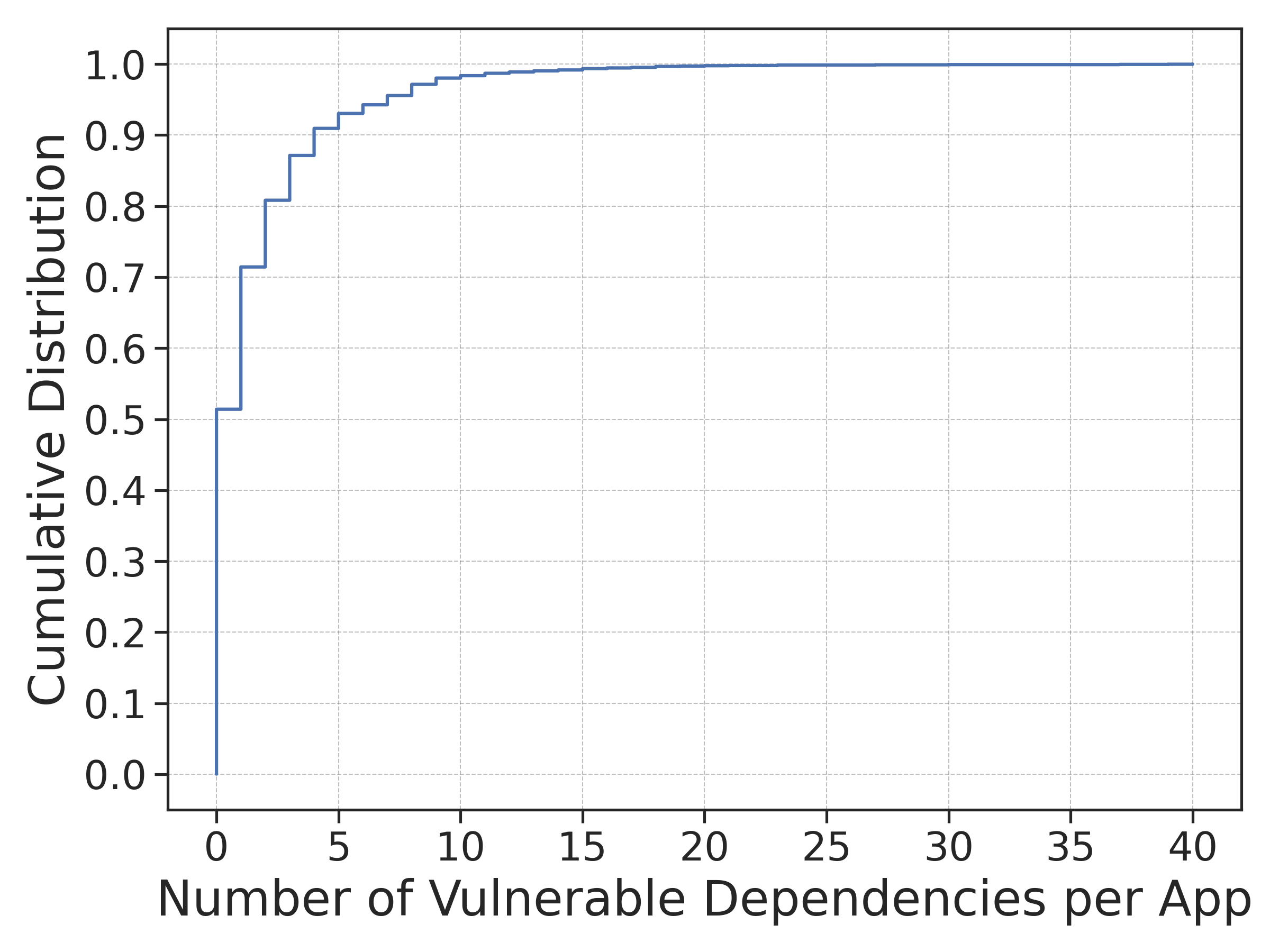}
        \caption{Cumulative Distribution of Vulnerable Dependencies per App}
        \label{fig:cdf_vuln_package}
    \end{subfigure}
    \hfill
    \begin{subfigure}[b]{0.45\columnwidth}
        \includegraphics[width=\linewidth]{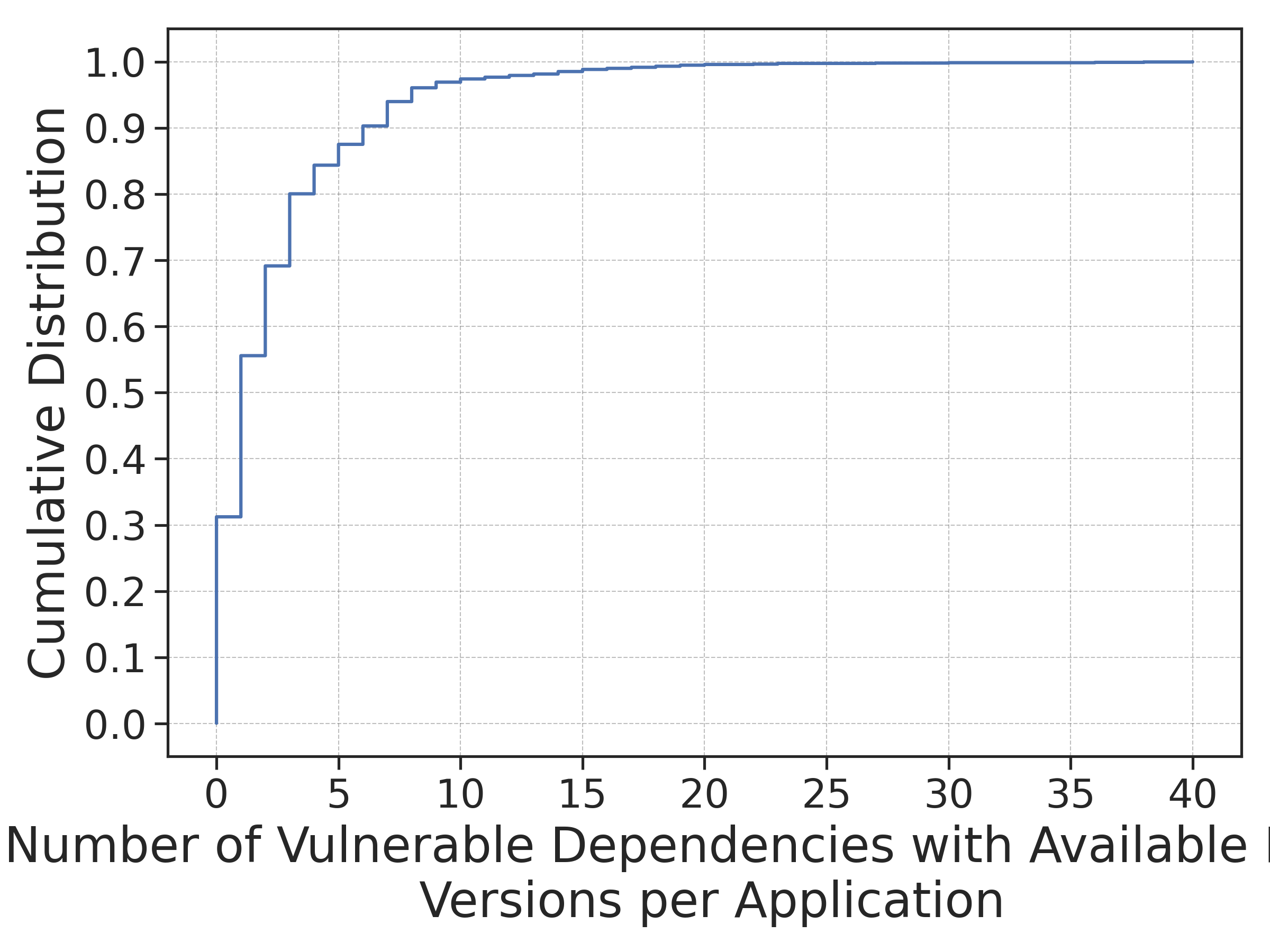}
        \caption{Cumulative Distribution of Dependencies with Fixed Version per App}
        \label{fig:cdf_fix_available}
    \end{subfigure}
    \caption{Cumulative distribution of dependencies across applications}
    \label{fig:cumulative_distribution}
\end{figure}

Then, we investigate the risks of applications from the vulnerable dependencies view. Figure~\ref{fig:cdf_vuln_package} shows the cumulative distribution of vulnerable dependencies in all applications collected. We observe that 51\% of applications are built without relying on any vulnerable dependencies, while the remaining 49\% depend on libraries with known vulnerabilities. Notably, some applications, such as \textit{OnurKerimoglu/facemoods}, \textit{ygtxr1997/ReliableSwap\_Demo}, and \textit{Honglee003/Background\_Removal\_and\_Change}, rely on more than 35 vulnerable dependencies, which exposes them to potentially higher security risks. These findings highlight the necessity for proactive dependency management to mitigate such risks in practice. 

To understand the potential to mitigate vulnerabilities, we analyze the distribution of patchable dependencies in applications with vulnerable dependencies, as shown in Figure~\ref{fig:cdf_fix_available}. While nearly 30\% of applications have no available fixed versions, 70\% can reduce security risks through timely updates. Specifically, \textbf{293} applications contain more than 5 dependencies with available fixed versions, indicating a greater potential for security improvement through proactive dependency management. This highlights the necessity for developers to keep their dependencies up-to-date to effectively mitigate known vulnerabilities.

Figure~\ref{fig:critical_CVE_count} presents the top 20 applications ranked by the number of their contained critical vulnerabilities. All these top 20 applications have more than 15 critical vulnerabilities, highlighting the crucial need for processes of managing the implementations of LLM-enabled applications. Among these applications, \textit{text-summarizer-for-news-articles} is the most vulnerable application, containing 33 critical vulnerabilities, which should be used carefully.

\begin{tcolorbox}[size=title,opacityfill=0.1]
\noindent
\textbf{Finding 9:} 49\% of the collected applications depend on libraries with known vulnerabilities, and 70\% of these applications rely on libraries for which fixed versions are already available. Moreover, \textit{text-summarizer-for-news-articles}, \textit{ReliableSwap\_Demo}, and \textit{visual\_chatgpt} represent the three most vulnerable applications that warrant careful attention. This underscores the necessity of inspecting the supply chain—specifically, assessing the influence of library dependencies on applications and addressing known vulnerabilities through timely fixes and dependency management.
\end{tcolorbox}

\section{Discussion}
\label{sec:discussion}

\subsection{Implication}

Based on our analysis, we provide implications for different stakeholders in LLMSC.

\textit{For application developers and managers.} With the significant growth of LLM-enabled applications, the risk assurance of these applications is important. Application developers should carefully use third-party libraries to avoid introducing potential risks to applications, such as \textit{pillow}, \textit{langchain}, and \textit{paddlepaddle}, which contain the most critical vulnerabilities. Application managers from the platform side~(e.g., Hugging Face) need to pay more attention to security and privacy guarantees in the uploaded applications, as these two types of risks are the most common ones. Moreover, we observe that, unlike traditional software libraries—which are often supported by well-maintained vulnerability databases—models and datasets currently lack such systematic tracking. This gap makes it challenging to perform thorough risk analysis and monitor known issues in these components. From an industry perspective, there is a clear need to establish dedicated vulnerability databases for models and datasets, similar to CVE systems for software libraries, to support transparent risk management and secure deployment of LLM-based systems.

\textit{For model developers and managers.} Before uploading models to the model database, model developers need to comprehensively evaluate their security and privacy properties. For example, model security and privacy should be addressed carefully.
For model managers on platforms, in addition to commonly used model scan tools, e.g., ModelScan, it is also necessary to employ more techniques to detect potential risks in models, e.g., hallucination detection methods~\cite{luo2024hallucination}. Moreover, model providers should disclose known issues to ensure that users are aware of potential risks and can implement appropriate mitigation strategies at the application level.

\textit{For dataset providers and managers.} Data cleaning is crucial, as most risks are caused by it. Dataset maintainers are recommended to employ tools to detect poisoned and privacy-sensitive data. As 28\% of the risks come from the root cause of \textit{Vulnerable Hugging Face Datasets Library}, managers need to focus on the implementation of libraries relevant to the datasets and remove such risks. A transparent dataset construction process is also essential, enabling users to understand dataset dependencies and trace potential supply chain risks associated with the data they rely on.

\textit{For LLMSC researchers.} During our collection and analysis process, we encountered several key challenges: 1) accurately identifying dependency relationships based on limited and often unstructured metadata; 2) extracting and validating risk issues for each component—particularly for models and datasets, where risk information is sparse or undocumented; and 3) demonstrating how known vulnerabilities propagate through dependencies to affect the overall security of the application.
These challenges highlight the need for the LLMSC research community to develop more advanced and automated toolchains that can support robust dependency analysis, systematic risk identification, and end-to-end vulnerability impact demonstration across the LLM supply chain.

\textit{Real-world Application Risk Mitigation.} We find that 49\% of the collected applications depend on libraries with known vulnerabilities, and nearly 70\% of them have dependencies with available fixed versions. This finding suggests that application developers should proactively check whether their applications depend on vulnerable libraries. Once identified, such risks can be effectively mitigated by regularly updating dependency libraries to their latest versions. As illustrated in~\cite{optimum-habana-issue1912}, this highlights the importance of timely updates in reducing security risks.

\subsection{Threat to Validity}

For \textit{internal validity}, the mapping of collected issues to our 13-factor risk taxonomy and the heuristic weighting of scores are constructed manually, which may introduce subjectivity. To mitigate this, we conduct sensitivity analyses and double-check manual inspection to confirm the stability of our results.

Regarding \textit{external validity}, our dataset is constrained to LLM applications available on Hugging Face, as we considered Hugging Face a comprehensive and representative platform for contemporary LLM applications. All applications are collected from a snapshot taken on 18 July 2025 within a three-year period, as our focus is on the most recent dataset within three years to ensure comprehensiveness and timeliness. As we only consider dependent models and datasets that have clear information provided on the platform, and the libraries listed in \textit{requirements} files, \dataset may not cover all dependencies. Moreover, since there is no ground truth for risks for applications, models, and datasets, risk issues in \dataset could have biases. However, \dataset is the first benchmark dataset that includes multi-components of LLMSC, which we believe its potential to facilitate future research in this domain. We plan to actively maintain \dataset and try to guarantee its comprehensiveness.

\section{Related Work}
\label{sec:related}
\subsection{Research on Security Concerns in LLMs}

Existing LLM security assurance works mainly focus on the model itself. Most studies employ jailbreak attack methods \cite{xu2024comprehensive,wei2023jailbroken,shen2024anything} or adversarial attack techniques \cite{carlini2023aligned,liu2024exploring,shayegani2023survey,zou2023universal} to assess the robustness of LLMs, aiming to induce unintended outputs or bypass safety filters. Other works \cite{li2024backdoorllm,zhao2024survey} have revealed that attackers can implant hidden backdoors in the model, allowing a compromised LLM to perform normally on benign samples but behave maliciously on specific trigger inputs. In addition, LLMs are also vulnerable to prompt injection \cite{liu2024automatic,liu2024formalizing,zhang2024goal} and model inference attacks, such as attribute inference~\cite{9152761,staab2023beyond} and membership inference attacks \cite{hu2025membership,das2025blind,duan2024membership}. Although these studies primarily focus on the risk and trustworthiness of LLMs themselves, our work conducts a comprehensive
investigation that accounts for the risks across all components (library, dataset, model, and application) within real-world applications.

\subsection{Research on LLM Applications}
Several studies have focused on analyzing the landscape and security issues of custom LLM applications. \cite{zhao2025llm} provides a vision and roadmap for LLM app store analysis, focusing on key aspects such as app data collection, security and privacy analysis, and ecosystem and market analysis. \cite{hou2024gptzoo} introduces GPTZoo, a large-scale dataset comprising 730,420 GPT instances to support academic research on GPTs. \cite{yan2024exploring} examines the distribution and deployment models in the integration of LLMs and third-party apps, and assesses their security and privacy implications. \cite{zhao2024gpts} analyzes the landscape of custom ChatGPT models, including community perception, GPT details, and GPT authors. \cite{su2024gpt} conducts a measurement study on the GPT Store, focusing on GPT categorization, popularity factors, and risks. Other works investigate security issues such as LLM app squatting and cloning~\cite{xie2024llm}, abusive or malicious behaviors~\cite{hou2025security}, misused by adversaries~\cite{antebi2024gpt}, and security and privacy in LLM platforms \cite{iqbal2024llm,tao2023opening}. However, these works mainly focus on security issues in custom LLM apps assembled through prompt engineering or plugin configuration, particularly those involving customized instructions, knowledge bases, and external plugins. In contrast, our analysis targets real-world LLM applications built through conventional software development workflows and conducts a comprehensive security analysis across datasets, libraries, and models.

Recent works have targeted security issues in real-world LLM applications. For example, \cite{wang2025sok} analyzes 529 vulnerabilities reported across 75 prominent projects to understand vulnerabilities in the LLMSC. \cite{shao2024llms} conducts a comprehensive study of 100 open-source applications that incorporate LLMs with RAG support and identifies 18 defect patterns. Other studies have revealed that such applications are vulnerable to prompt leakage \cite{su2024gpt,hui2024pleak}, prompt injection \cite{liu2023prompt,liu2310prompt,pedro2023prompt} and security issues arising from the integration of LLMs with other components, such as RAG~\cite{shao2024llms,wu2024new}. However, these works mainly focus on security issues within particular components. In contrast, our work systematically investigates security issues, thereby providing a more comprehensive understanding of risks within LLM-integrated applications.

\section{Conclusion}

In this paper, we introduce the first dataset, \dataset, to support the risk assurance of LLMSC. \dataset covers a comprehensive information of 3,859 LLM-enabled applications, 109,211 LLMs, 2,474 datasets, and 8,862 implementation libraries, and provides detailed risk issues hidden in each LLMSC component. Via a systematic analysis of \dataset, we reveal multiple findings relevant to LLMSC. Moreover, based on the analysis, we provide insightful implications for different stakeholders. We hope this research and our dataset can facilitate the risk assurance of LLMSC. 